\PassOptionsToPackage{hyphens}{url}
\documentclass[sigconf, screen, nonacm]{acmart}

\setcopyright{none}
\acmConference[]{}{}{}
\acmYear{2026}
\acmISBN{}
\acmDOI{}
\renewcommand\footnotetextcopyrightpermission[1]{}

\usepackage[ruled,vlined]{algorithm2e}
\usepackage{textcomp}
\usepackage{tikz}
\usepackage{pgfplots}
\pgfplotsset{compat=1.18}
\usetikzlibrary{plotmarks}
\usetikzlibrary{positioning, shapes.geometric, patterns, calc}

\usepackage[many]{tcolorbox}
\usepackage{fancyvrb}
\usepackage{pifont}
\usepackage{float}
\newfloat{listing}{tbp}{lol}
\floatname{listing}{Listing}
\definecolor{findrule}{HTML}{2C6E9B}   % steel blue: label
\definecolor{findback}{HTML}{E8F3FB}   % light blue panel
\definecolor{implback}{HTML}{E8F6EA}   % light green panel
\definecolor{impldark}{HTML}{1B5E20}   % dark green label
\definecolor{codecmt}{HTML}{388E3C}    % light green: code comments
\newcounter{finding}
\newtcolorbox{findingbox}{
  enhanced, nobeforeafter=false,
  colback=findback, colframe=black, boxrule=0.6pt, arc=8pt,
  drop fuzzy shadow={black!30},
  left=7pt, right=6pt, top=6pt, bottom=6pt,
  before skip=7pt plus 2pt, after skip=7pt plus 2pt}
\newtcolorbox{implbox}{
  enhanced, nobeforeafter=false,
  colback=implback, colframe=black, boxrule=0.6pt, arc=8pt,
  drop fuzzy shadow={black!30},
  left=7pt, right=6pt, top=6pt, bottom=6pt,
  before skip=7pt plus 2pt, after skip=7pt plus 2pt}
\newenvironment{finding}
  {\refstepcounter{finding}\begin{findingbox}%
   \textbf{\textcolor{findrule}{Finding \thefinding.}}\hspace{0.4em}\ignorespaces}
  {\end{findingbox}}

\newcounter{implication}
\newenvironment{implication}
  {\refstepcounter{implication}\begin{implbox}%
   \textbf{\textcolor{impldark}{Implication \theimplication.}}\hspace{0.4em}\ignorespaces}
  {\end{implbox}}

\begin{document}

\title{Dissecting the GPU Load/Store Unit and Exposing Its Security Risks}
\author{Wenbo Han}
\affiliation{%
  \institution{University of Luxembourg}
  \country{Luxembourg}
}
\email{wenbo.han@uni.lu}

\author{Pierre Talbot}
\affiliation{%
  \institution{University of Luxembourg}
  \country{Luxembourg}
}
\email{pierre.talbot@uni.lu}

\renewcommand{\shortauthors}{Han and Talbot}

\begin{abstract}
GPUs are the main platform for large-scale machine learning and power the largest supercomputers.
Their load/store units (LSUs) often determine kernel performance and have been exploited in prior
attacks.
Yet what is known about the LSU is only a few isolated mechanisms and parameters, and even a single
one has been enough for an attack.
In particular, it is still unknown what a load costs, where loads contend, and even whether a warp
can starve due to contention.
Until each is measured, neither its security risk nor a defense is known.
We reverse-engineer how the LSU serves load instructions on recent NVIDIA GPUs (V100, A100, H100) with microbenchmarks
and analyze the security implications.
We give a cycle-accurate latency formula for a global L1 hit or shared memory access from the
addresses it touches, whose address-dependent terms are bank conflicts and the L1 tag lookup; we
show that loads contend at both, and that when the LSU pipeline is full a warp can starve a warp of
a later-launched block.
We then show that the LSU's attack surface is far broader than prior attacks revealed:
through channels never exploited before, it can leak information, communicate covertly, and
degrade colocated workloads.
Through either global bank conflicts or tag lookups alone, we recover a full AES-128 key under the
threat model of prior GPU AES timing attacks.
Under NVIDIA's Multi-Process Service (MPS), the bank access and the tag lookup each open a covert
channel across processes, whose rate rises with each GPU generation.
Also under MPS, this starvation lets one warp slow a colocated cuBLAS matrix multiplication by up to
750$\times$ on the A100 and 563$\times$ on the H100.
\end{abstract}


\maketitle

\section{Introduction}
\label{sec:intro}

GPUs have become the main platform for large-scale machine learning and power the largest
supercomputers: the
training and inference of large language models run on clusters of thousands of
GPUs~\cite{megatron21}, and eight of the ten fastest supercomputers are built
around GPUs~\cite{top500jun26}.
The high performance of GPUs comes from the high throughput of massive parallelism: NVIDIA GPUs, which this paper studies, consist of tens to over a hundred streaming multiprocessors (SMs), each of which keeps
thousands of threads resident.
This throughput depends on how fast these threads access memory: when memory returns data fast
enough, the GPU can fully exploit its parallelism.
Yet how long a single memory instruction takes is not well understood, and this timing also
creates security risks~\cite{trident21,jiang16,jiang19}.

Almost every load and store instruction a kernel issues passes through the load/store unit (LSU)
of its SM, and the design of the LSU determines part of its latency. NVIDIA treats this unit as a central performance concern: the CUDA Best Practices Guide devotes a
section to shared-memory bank conflicts~\cite{cudabp}, and several of Nsight Compute's stall
metrics are attributed to it~\cite{ncuguide}. Many high-performance kernels are built around it.
NVIDIA's CUTLASS general matrix multiplication (GEMM), for instance, stages its operands through shared-memory tiles~\cite{cutlass},
and FlashAttention streams the key and value matrices of attention through the same shared memory,
reusing them to cut memory traffic~\cite{dao22,dao23,shah24}. In current designs, the LSU makes kernels fast
partly because one on-chip SRAM per SM, which holds shared memory and caches global data, lets
warps reuse one another's loads~\cite{v100wp,a100wp,h100wp}. For the same reason, almost every
memory instruction on the SM competes for the LSU. And because the latency of each
access depends on the addresses it touches, the unit can also serve as a side channel.
Indeed, parts of its mechanisms and parameters have already been exploited: the linear relation between execution
time and the number of cache lines a load touches and the hit/miss timing of the
L1 cache leak AES keys~\cite{trident21,jiang16}, and shared-memory bank conflicts leak them as
well~\cite{jiang19}. These timing channels are serious enough that defenses have been
proposed~\cite{kadam18,kadam20,xu19}.

Yet what is known about this unit is a set of isolated mechanisms and parameters.
The latency of a single load has been measured, but without regard to the addresses it touches or
to how they interact with the access width.
The cost of a bank conflict is known only approximately, as about two extra cycles per additional
way averaged over many shared loads~\cite{mei16,jia18,jia19,sun22,jiang19,huerta25}. Huerta et al.\ measured when loads issue and how long a single load takes,
without varying the addresses they touch~\cite{huerta25}. As a result, given the addresses a load touches, prior work cannot tell how many cycles the LSU takes. In particular, the following questions
remain open: (1) when a global load hits in the L1 cache, which of its cache lines
cost extra cycles; (2) whether each additional way of bank conflict costs exactly two cycles for
a shared load; (3) whether a global load pays the same for a bank conflict as a shared load;
(4) whether, combining these factors, one can give the exact number of cycles a load takes;
(5) how many loads the LSU serves at once; and (6) whether a warp can starve when it is full. Even a single mechanism or parameter
has been enough for an attack~\cite{trident21,jiang16,jiang19}; thus, what remains undiscovered is likely
to hide more.

In this paper, we answer these questions by reverse-engineering how the LSU serves loads
and analyzing the security implications. The answers matter because the coverage of a defense can
only be judged against the channels that are known; what the unit does beyond them is untested, not
safe.

To answer them, we mainly design timing microbenchmarks in SASS, the assembly language of NVIDIA
GPUs, that focus on the LSU. Timing a short SASS sequence with the SM's cycle counter is an
established method~\cite{zhang17,jia18,raihan19}, and it is how Huerta et al.\ measured the cycle at
which each load issues~\cite{huerta25}. However, prior work varied the instructions in the timed sequence, never the addresses that the
lanes of a load touch. By assigning each lane of a load a chosen address and timing the load, we can infer what happens inside the LSU even though no instruction
exposes it directly. We expose how many cycles each step takes, and thus where the cycles of a load
go from its issue to its completion, together with \emph{the exact number of cycles a single load takes}, for a shared load or a
global load that hits in the L1 cache.
If we also record when each load issues and completes, we obtain a timeline of a sequence of loads.
From it we can read how many loads a unit serves at once.

This work makes the following contributions:

\begin{itemize}
\item On the V100, A100, and H100, we reverse-engineer how the LSU serves loads. One key result is a cycle-accurate formula that predicts a load's latency from the addresses it touches, for shared loads on all three GPUs and for global loads that hit in the L1 cache on the A100 and H100: a base cost, plus a tag-lookup cost for global loads, plus a bank-access cost.

\item We find that at most two instructions access the banks at the same time, whereas the four tag-lookup units serve only one instruction at a time for the whole SM.

\item We find that within a sub-core, when the LSU pipeline is full, the loads of a warp of an earlier-launched block issue before those of a later one, so a warp that always keeps a load ready can starve a warp of a block launched after it.

\item Under NVIDIA's Multi-Process Service (MPS), which lets kernels of different processes run concurrently on the same GPU, we show one warp can use this starvation to degrade 11 NVIDIA library kernels (from cuBLAS, cuFFT, cuSPARSE, CUB, and Thrust) that share its SM, slowing a cuBLAS GEMM by up to 563$\times$ on the H100.

\item Also under MPS, we build a cross-process covert channel from contention for the bank-access channels and the tag-lookup units, with a rate that rises from $20.0$\,Mb/s on the V100 to $44.0$\,Mb/s on the H100.

\item We demonstrate the existence of two side channels in GPUs, timing the kernel directly as in prior work's main experiments~\cite{jiang16,trident21,jiang19}, and recover a full AES-128 key from the cost of a tag lookup and of a global load's bank conflicts.
\end{itemize}

\section{Background}
\label{sec:bg}

\subsection{GPU Architecture}
\label{sec:arch}

\begin{figure}[t]
\centering
\includegraphics[width=0.85\columnwidth]{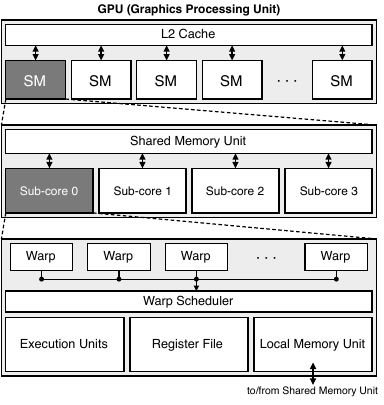}
\caption{Modern NVIDIA GPU architecture.}
\label{fig:gpu-arch}
\end{figure}

Fig.~\ref{fig:gpu-arch} shows the typical architecture of a modern NVIDIA GPU. The GPU consists
of many streaming multiprocessors (SMs) that share an L2 cache. Each SM runs many threads at
once and is divided into four sub-cores, each with its own warp scheduler, register file, and
execution units~\cite{v100wp}. Threads are grouped into warps of 32 under the
single-instruction, multiple-thread (SIMT) model: the active threads of a warp execute the same
instruction together, so one load instruction accesses up to 32 addresses, one per
thread~\cite{v100wp}.

Memory instructions are executed by the load/store unit (LSU), which has two parts. The first is
a local memory unit in each sub-core, in which, as Huerta et al.\ infer, the addresses are
calculated~\cite{huerta25}. The second, the shared memory unit in Fig.~\ref{fig:gpu-arch}, is
shared by the four sub-cores because it holds the SM's one on-chip SRAM. This SRAM holds both
memory spaces a load can access: global memory is cached in the L1 data cache, and shared memory
is used by the threads of a block to exchange data~\cite{v100wp,a100wp}. In SASS, a load from
global memory is the instruction \texttt{LDG} and a load from shared memory is \texttt{LDS}. The access itself
therefore happens in this shared unit.

How the 32 addresses of a load spread over the banks of the SRAM decides how the load is
served. The SRAM is divided into 32 banks, and consecutive 4-byte words fall into consecutive
banks~\cite{cudabp}. A bank conflict occurs when the threads of one load access different
words in the same bank~\cite{aamodt18,cudabp}. The hardware then serializes these accesses,
splitting the load into several passes~\cite{cudabp}, and the cost is set by the busiest
bank~\cite{jiang19}. We write $\mathit{bf}$ for the number of different words in the
busiest bank; a load with $\mathit{bf}=n$ thus has an $n$-way bank conflict. NVIDIA's guides
and prior work discuss bank conflicts only for shared memory~\cite{cudabp,jiang19,sun22}, but
global loads read the same SRAM, and Nsight Compute reports bank conflicts for them as
well~\cite{ncuguide}. Before it reads the L1 cache, a global load must, in addition, determine
whether its data are there: its addresses fall into up to 32 cache lines of 128 bytes, and it looks up the
tag of each line to check whether the line is in the cache~\cite{nvforum25tag}.

\subsection{Dependency Resolution with Control Bits}
\label{sec:ctrlbits}

Each sub-core of a modern NVIDIA GPU can issue one instruction per cycle. In
every cycle, its warp scheduler selects one eligible warp and issues that
warp's next instruction in program order. One condition for a warp to be
eligible is that the dependencies of its next instruction have been resolved,
for example a read-after-write (RAW) dependency on a register that an earlier
instruction writes (Fig.~\ref{fig:deps}), or a write-after-read (WAR)
dependency on a register that an earlier instruction still has to read.

\begin{figure}[H]
\centering\scriptsize
\begin{tcolorbox}[enhanced, width=0.47\columnwidth, nobeforeafter, box align=top, equal height group=dep, colback=white, colframe=black, boxrule=0.6pt, arc=0pt, left=3pt, right=3pt, top=1pt, bottom=1pt]
\begin{verbatim}
FADD R1, R2, R3
MOV  R4, R1
\end{verbatim}
\end{tcolorbox}\hfill
\begin{tcolorbox}[enhanced, width=0.47\columnwidth, nobeforeafter, box align=top, equal height group=dep, colback=white, colframe=black, boxrule=0.6pt, arc=0pt, left=3pt, right=3pt, top=1pt, bottom=1pt]
\begin{verbatim}
LDG  R1, [R2]
MOV  R4, R1
\end{verbatim}
\end{tcolorbox}\\[2pt]
\begin{minipage}[t]{0.47\columnwidth}\centering (a)\end{minipage}\hfill
\begin{minipage}[t]{0.47\columnwidth}\centering (b)\end{minipage}
\caption{Dependencies in two snippets. In both, the \texttt{MOV} instruction reads \texttt{R1} (RAW), which (a) a fixed-latency \texttt{FADD} instruction or (b) a variable-latency \texttt{LDG} instruction writes.}
\label{fig:deps}
\end{figure}

Modern NVIDIA GPUs do not detect these dependencies with a hardware scoreboard
that compares register names. Instead, the compiler analyzes the dependencies
and encodes control bits into the binary encoding of each
instruction~\cite{jia18,jia19,huerta25,cuassembler}. When an instruction is
about to issue, and when it issues, the hardware does exactly what its control
bits specify.

If the earlier instruction has a fixed latency, like the \texttt{FADD} instruction in Fig.~\ref{fig:deps}(a), the compiler encodes, in the binary encoding of this instruction, a value
for the warp's stall counter (SC), long enough for the result to be ready. When the instruction
issues, the SC is set to this value; it then decreases by one every cycle, and
the warp becomes eligible again only when it reaches zero. An instruction that
sets $\mathrm{SC}=1$ thus allows the next instruction to issue in the next cycle.

If its latency varies, like that of the \texttt{LDG} instruction in Fig.~\ref{fig:deps}(b), the compiler cannot know the latency at compile time, so no SC fits.
The hardware therefore provides dependence counters (DCs), six per warp. An instruction $I$ can name one of them as its write DC and another as its read DC, when
later instructions depend on its destination register or on its source registers,
respectively. Either DC is incremented when $I$ issues and decremented, which we call
\emph{released}, when $I$ has written its destination register (write DC) or has read its
source registers (read DC). The compiler likewise encodes, in the control bits of each
instruction that depends on $I$, which DCs must be back to zero before it can issue. For
example, the \texttt{MOV} instruction in Fig.~\ref{fig:deps}(b) depends on the \texttt{LDG} instruction through
\texttt{R1} (RAW). The compiler therefore names a write DC for the \texttt{LDG} instruction, say DC1, and lets the \texttt{MOV} instruction wait on DC1, so it issues only after the loaded value is
available.

\subsection{GPU Microarchitectural Security}
\label{sec:security}

As GPUs take on security-sensitive and multi-tenant workloads, their
microarchitecture has become the target of a wide range of attacks. We review the
three classes relevant to this work: performance degradation, covert channels, and
side channels.

\paragraph{Performance degradation}
A performance-degradation attack makes a victim's workload run slower; on GPUs this is
most often achieved by contending for a shared
resource~\cite{bechtel22,olmedo20,yandrofski22,wang25}. For
example, Wang et al.\ have an attack kernel pin useless lines in every L2 set with
\texttt{evict\_last} eviction hints; this shrinks the capacity left for the victim
kernel, pushing much of its working set out to DRAM so that many of its accesses
become slow DRAM fetches~\cite{wang25}.

\paragraph{Covert channels}
A covert channel transfers information between a sender and a receiver through a medium
not meant for communication; like a degradation attack, it is most often built on
contention for a shared resource~\cite{naghib17,ahnnoc21,tunnels23,dutta23,mc3,wang25}.
Typically, building one comes down to two questions: (1) \emph{how to encode a single
bit}, and (2) \emph{how to keep the sender and receiver synchronized so that a long
stream of such bits arrives in order}. For example, contending for one cache set to carry a bit is a mainstream
covert channel on both CPUs~\cite{llc15} and GPUs~\cite{naghib17}: the receiver first
fills the set with its own cache lines; to send a ``1'' the sender evicts the whole set,
and to send a ``0'' it does nothing; the receiver then re-reads its lines and tells the
two apart by the latency---slow means evicted (a ``1''), fast means still cached (a
``0'').

\paragraph{Side channels}
A timing side channel arises when a program's running time depends on the secret it
processes, so that an attacker who can obtain the timing, together with known or chosen
inputs or the resulting ciphertext, can infer part or all of the secret. Such an attack normally needs three ingredients:
(1) \emph{an implementation whose work depends on the secret}; (2) \emph{a
microarchitectural effect that turns that dependence into something observable}; and
(3) \emph{a method that measures the effect and analyzes it to recover the key}.

AES is a standard block cipher, and AES-128 encrypts each block in 10 rounds. On a
GPU, each encryption block is typically handled by one thread. A normal implementation
computes its last round with T4-table lookups indexed by key-dependent bytes~\cite{iwai10,tezcan21}.
These key-dependent lookups have repeatedly been exploited to recover full AES-128 keys on GPUs~\cite{jiang16,trident21,jiang19}.
For example, Jiang et al.~\cite{jiang16} exploit SIMT leakage---a load's time grows with the number of distinct cache lines
the warp touches. The last-round table index can be computed from the ciphertext and a
guessed key byte, so for each guess from 0 to 255 the attacker predicts how many
distinct lines the lookup would touch and correlates this prediction with the measured
time; only the correct guess predicts the real line count, so only it correlates, and
the byte with the strongest correlation is the key. On today's sectored caches this
positive correlation alone no longer recovers the key; Trident~\cite{trident21} combines
it with the timing difference between L1 hits and misses to recover the full key. Jiang
et al.~\cite{jiang19} use the linear relation between shared-memory bank conflicts and
time to recover the key when the table sits in shared memory.

\section{Methodology}
\label{sec:method}
\label{sec:cycletable}

\begin{figure}[t]
\centering\scriptsize
\begin{minipage}[c]{0.53\columnwidth}
\begin{tcolorbox}[enhanced, width=\linewidth, nobeforeafter, colback=white, colframe=black, boxrule=0.6pt, arc=0pt, left=3pt, right=3pt, top=1pt, bottom=1pt]
\begin{Verbatim}[commandchars=\\\{\}]
CS2R R2, SR_CLOCKLO  \textcolor{codecmt}{# SC=1}
LDS  R1, [R0]        \textcolor{codecmt}{# SC=2, sets DC1}
CS2R R3, SR_CLOCKLO  \textcolor{codecmt}{# waits on DC1}
\end{Verbatim}
\end{tcolorbox}
\centering (a)\\[4pt]
\begin{tcolorbox}[enhanced, width=\linewidth, nobeforeafter, colback=white, colframe=black, boxrule=0.6pt, arc=0pt, left=3pt, right=3pt, top=1pt, bottom=1pt]
\begin{Verbatim}[commandchars=\\\{\}]
CS2R R2, SR_CLOCKLO  \textcolor{codecmt}{# SC=1}
LD   R4, [R0]        \textcolor{codecmt}{# SC=1}
LD   R6, [R0]        \textcolor{codecmt}{# SC=1}
...
\textcolor{codecmt}{# issue: SC=1; release: SC=2, sets DC1}
LDi  RN, [R0]
\textcolor{codecmt}{# issue: no wait; release: waits on DC1}
CS2R R3, SR_CLOCKLO
\end{Verbatim}
\end{tcolorbox}
\centering (b)
\end{minipage}\hfill
\begin{minipage}[c]{0.44\columnwidth}
\centering
\begin{tikzpicture}[
  x=1cm,y=1cm,
  cmp/.style={circle,fill=black,inner sep=0pt,minimum size=2.7pt},
  iss/.style={circle,draw=black,fill=white,line width=.5pt,inner sep=0pt,minimum size=2.7pt},
  lane/.style={draw=black!45,line width=.5pt},
  nolane/.style={draw=black!15,line width=.5pt},
  dim/.style={draw=black!70,line width=.5pt},
  rul/.style={draw=black!60,line width=.45pt},
  band/.style={fill=black!5},
  lab/.style={font=\small},
  tiny/.style={font=\footnotesize,color=black!75}]
\providecommand{\brk}[2]{% ... in place of the omitted stretch
  \foreach \dx in {-0.09,0,0.09}{\fill[black!55] (#1+\dx,#2) circle (0.55pt);}}
\node[lab,anchor=east] at (-0.34,-0.694) {\texttt{LDG}};
\fill[band] (0.000,-0.246) rectangle (1.404,-1.067);
\draw[draw=black!12,line width=.3pt] (0.000,-0.246) -- (0.000,-1.067);
\draw[draw=black!12,line width=.3pt] (1.404,-0.246) -- (1.404,-1.067);
\draw[lane] (0.000,-0.694) -- (2.808,-0.694);
\draw[line width=.65pt] (0.000,-0.694) -- (0.000,-0.422); \node[cmp] at (0.000,-0.389) {};
\draw[line width=.65pt] (0.351,-0.694) -- (0.351,-0.966); \node[iss] at (0.351,-0.999) {};
\draw[line width=.65pt] (1.404,-0.694) -- (1.404,-0.422); \node[cmp] at (1.404,-0.389) {};
\draw[line width=.65pt] (1.755,-0.694) -- (1.755,-0.966); \node[iss] at (1.755,-0.999) {};
\draw[line width=.65pt] (2.808,-0.694) -- (2.808,-0.422); \node[cmp] at (2.808,-0.389) {};
\draw[dim] (0,-0.366) -- (0,-0.316) -- (1.404,-0.316) -- (1.404,-0.366);
\node[tiny,anchor=south west,inner sep=.5pt] at (-0.020,-0.311) {one period $=4$ cycles};
\node[lab,anchor=east] at (-0.34,-1.814) {\texttt{LDS}};
\fill[band] (0.000,-1.434) rectangle (0.702,-2.114);
\draw[draw=black!12,line width=.3pt] (0.000,-1.434) -- (0.000,-2.114);
\draw[draw=black!12,line width=.3pt] (0.702,-1.434) -- (0.702,-2.114);
\fill[band] (1.404,-1.434) rectangle (2.106,-2.114);
\draw[draw=black!12,line width=.3pt] (1.404,-1.434) -- (1.404,-2.114);
\draw[draw=black!12,line width=.3pt] (2.106,-1.434) -- (2.106,-2.114);
\draw[lane] (0.000,-1.814) -- (2.808,-1.814);
\draw[line width=.65pt] (0.000,-1.814) -- (0.000,-1.644); \node[cmp] at (0.000,-1.611) {};
\draw[line width=.65pt] (0.000,-1.814) -- (0.000,-1.984); \node[iss] at (0.000,-2.017) {};
\draw[line width=.65pt] (0.702,-1.814) -- (0.702,-1.644); \node[cmp] at (0.702,-1.611) {};
\draw[line width=.65pt] (0.702,-1.814) -- (0.702,-1.984); \node[iss] at (0.702,-2.017) {};
\draw[line width=.65pt] (1.404,-1.814) -- (1.404,-1.644); \node[cmp] at (1.404,-1.611) {};
\draw[line width=.65pt] (1.404,-1.814) -- (1.404,-1.984); \node[iss] at (1.404,-2.017) {};
\draw[line width=.65pt] (2.106,-1.814) -- (2.106,-1.644); \node[cmp] at (2.106,-1.611) {};
\draw[line width=.65pt] (2.106,-1.814) -- (2.106,-1.984); \node[iss] at (2.106,-2.017) {};
\draw[line width=.65pt] (2.808,-1.814) -- (2.808,-1.644); \node[cmp] at (2.808,-1.611) {};
\draw[line width=.65pt] (2.808,-1.814) -- (2.808,-1.984); \node[iss] at (2.808,-2.017) {};
\draw[dim] (0,-1.554) -- (0,-1.504) -- (0.702,-1.504) -- (0.702,-1.554);
\node[tiny,anchor=south west,inner sep=.5pt] at (-0.020,-1.499) {one period $=2$ cycles};
\draw[rul] (0.000,-2.264) -- (2.808,-2.264);
\draw[rul] (0.000,-2.264) -- (0.000,-2.179);
\draw[rul] (0.351,-2.264) -- (0.351,-2.219);
\draw[rul] (0.702,-2.264) -- (0.702,-2.219);
\draw[rul] (1.053,-2.264) -- (1.053,-2.219);
\draw[rul] (1.404,-2.264) -- (1.404,-2.219);
\draw[rul] (1.755,-2.264) -- (1.755,-2.179);
\draw[rul] (2.106,-2.264) -- (2.106,-2.219);
\draw[rul] (2.457,-2.264) -- (2.457,-2.219);
\draw[rul] (2.808,-2.264) -- (2.808,-2.219);
\node[tiny,anchor=base west] at (0,-2.690) {1 tick $=$ 1 cycle};
\draw[->,black,line width=.5pt] (0,-2.764) -- (2.808,-2.764);
\node[anchor=base east,font=\footnotesize,color=black!70] at (2.808,-2.690) {time};
\node[cmp,anchor=west] (cr) at (0,-3.044) {};
\node[tiny,anchor=west,inner sep=.5pt] (lr) at ([xshift=1.5pt]cr.east) {release};
\node[iss,anchor=west] (ci) at ([xshift=6pt]lr.east) {};
\node[tiny,anchor=west,inner sep=.5pt] (li) at ([xshift=1.5pt]ci.east) {issue};
\brk{-0.17}{-0.694}
\brk{2.95}{-0.694}
\brk{-0.17}{-1.814}
\brk{2.95}{-1.814}
\brk{-0.17}{-2.264}
\brk{2.95}{-2.264}
\end{tikzpicture}\\
(c)
\end{minipage}
\caption{Snippets and readings. (a) Code to time one load.
(b) Code to read the issue or release cycle of load $i$; \texttt{LD} is \texttt{LDS} or \texttt{LDG}.
(c) Issue and release cycles of (b) on the A100 once the stream reaches steady
state; one tick is one cycle.}
\label{fig:snippets}
\end{figure}
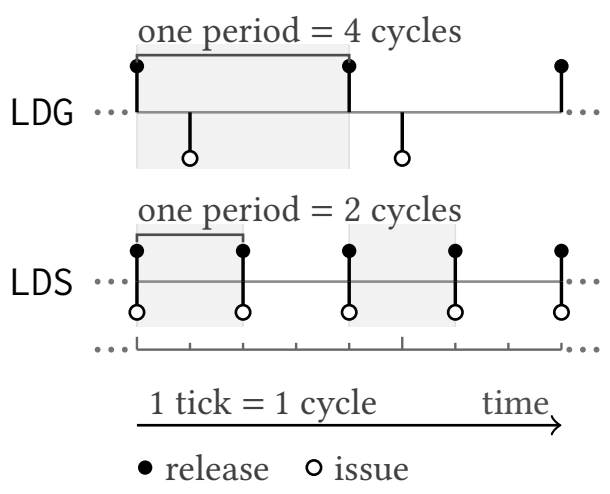

This section presents our timing microbenchmarks. We mainly use the two SASS snippets of
Fig.~\ref{fig:snippets} to reverse-engineer the LSU: snippet (a) gives the latency of one
instruction under a condition we choose, for example a given number of bank conflicts, and
snippet (b) builds the timeline of a sequence of loads described in \S\ref{sec:intro}.
Unless stated otherwise, every experiment runs
a single block of one warp (32 threads) on the V100, A100, and H100, and every load accesses 4 bytes per
thread. We edit the
compiler-generated SASS with CuAssembler~\cite{huerta25} so that
the snippet contains only the instructions we are interested in, with the
control bits we set. Since CuAssembler does not support the H100, we extended it to the
SASS of the H100 for the instructions our experiments need. Snippets that time an interval
read the SM cycle counter with two \texttt{CS2R} instructions, a start and a stop, and write the difference back to global memory; \texttt{CS2R} is the SASS instruction that reads the current clock into a register. By interpreting this difference in different ways, we
analyze the details of the microarchitecture.

The first reading takes the difference as a latency. Fig.~\ref{fig:snippets}(a) times one
load: a single \texttt{LDS} instruction sits between the two \texttt{CS2R} instructions, and the stop \texttt{CS2R} instruction waits on the DC of the \texttt{LDS} instruction. The figure is annotated with the SC and DC of every instruction. The \texttt{LDS} instruction sets SC\,=\,2 rather than 1 because a
DC takes effect only one cycle after the producer instruction that sets it issues, so a producer
instruction whose consumer instruction is the next instruction must set SC\,=\,2~\cite{huerta25}. The
difference is thus the latency of the load. For example, if we run the snippet repeatedly,
each run with one more way of bank conflict in the addresses in \texttt{R0}, we can tell
whether a bank conflict has a fixed cost, and how large it is (\S\ref{sec:bank}). Likewise, we
can change other conditions: the access width, a global load in place of the \texttt{LDS} instruction,
or how the addresses spread over cache lines.

To build the timeline of a sequence of loads, we need two instants for each load: the cycle at
which it issues, and the cycle at which it releases its DC, so that an instruction waiting on that DC may issue in the next cycle (\S\ref{sec:ctrlbits}). We call them the \emph{issue} and
\emph{release} cycles. Both can be read from the same difference between the two \texttt{CS2R}
values, which we denote $D$.

To make this precise, we take the start \texttt{CS2R} instruction to issue at cycle 1 as the baseline; the stop \texttt{CS2R} instruction therefore issues at cycle $D{+}1$.

\emph{Issue cycle.} To measure the issue cycle of an instruction $i$ in a
snippet, we truncate the snippet right after $i$ and add a stop \texttt{CS2R} instruction that waits on nothing. The stop \texttt{CS2R} instruction then issues right after
$i$, so $D$ is the issue cycle of $i$. For this, we set the SC of every instruction to 1 and clear every DC, so that no instruction waits for
another.

\emph{Release cycle.} Issue cycles alone, however, cannot tell how the
pipeline stages of the LSU serve the loads, for example whether two
loads are served at the same time or one after the other. We therefore also
read release cycles. To read the release cycle of a variable-latency
instruction that writes a register, we again truncate the snippet right after it, assign it a write DC, and add a stop \texttt{CS2R} instruction that waits on that DC. Since $D{+}1$ is the earliest cycle at which an instruction waiting
on that DC can issue, we define $D$ as the release cycle of the instruction. The snippet of
Fig.~\ref{fig:snippets}(a) has this form, so its reading is also the release cycle of its \texttt{LDS} instruction.

We apply both readings to a sequence of independent loads
(Fig.~\ref{fig:snippets}(b)); to measure load $i$, we truncate the sequence
right after it. Fig.~\ref{fig:snippets}(c) shows the resulting timeline for conflict-free
loads in steady state: each filled dot marks a release and each open dot an issue, and
shaded and white bands alternate to mark consecutive periods, the time after which the
pattern repeats. The readings reproduce the rate at which global loads issue, once every
four cycles, which Huerta et al.\ explain by the address calculation in each
sub-core~\cite{huerta25}. We further observe that shared loads issue once every two cycles,
and that each global load issues one cycle after a release, whereas a shared load issues in
the same cycle as a release. \S\ref{sec:chan} builds more complex timelines than
Fig.~\ref{fig:snippets}(c), with more variables such as the bank conflicts of the loads and
the number of warps; from them we can read off how many loads issue and release in one period.

\section{How the LSU Serves Loads}
\label{sec:machine}

In this section, we reverse-engineer how the LSUs of modern NVIDIA GPUs serve loads, using the
methodology described in \S\ref{sec:method}. We start with the cost of a bank conflict (\S\ref{sec:bank}), which is at the heart of every experiment presented next. We then
show how the GPU looks up the tags of a load (\S\ref{sec:tag}) and what this lookup costs (\S\ref{sec:tagcost}). From these costs, we derive how many cycles a load instruction takes in total (\S\ref{sec:formula}). These are the costs of one load; we next measure how many loads the banks
(\S\ref{sec:chan}) and the tag-lookup units (\S\ref{sec:tagserial}) serve at once. Finally, we fill
the LSU pipeline and ask which load issues when it is full, which decides whether a warp can starve
(\S\ref{sec:buffer}).

\subsection{The Cost of Accessing the Banks}
\label{sec:bank}

\begin{figure}[t]
\centering
\resizebox{\columnwidth}{!}{\begin{tikzpicture}
\begin{axis}[at={(0,0)}, scale only axis, width=2.1cm, height=2.3cm,
  xmin=0, xmax=33, ymin=15, ymax=120, xtick={1,8,16,24,32}, minor xtick={1,...,32},
  ytick={20,40,...,120}, minor ytick={20,25,...,120},
  major tick length=3pt, minor tick length=1.4pt, tick align=outside, tick pos=left,
  tick label style={font=\scriptsize}, label style={font=\scriptsize},
  xlabel={$\mathit{bf}$}, xlabel style={yshift=1ex},
  title={(a) A100}, title style={font=\scriptsize, yshift=-1ex},
  legend style={font=\scriptsize, draw=black!40, fill=white, at={(0.04,0.97)}, anchor=north west},
  ylabel={latency (cycles)}, ylabel style={yshift=-1ex},
]
\addplot[blue!80!black, mark=square*, mark size=1.2pt, line width=.6pt] coordinates {(1,24) (2,26) (3,28) (4,30) (5,32) (6,34) (7,36) (8,38) (9,40) (10,42) (11,44) (12,46) (13,48) (14,50) (15,52) (16,54) (17,56) (18,58) (19,60) (20,62) (21,64) (22,66) (23,68) (24,70) (25,72) (26,74) (27,76) (28,78) (29,80) (30,82) (31,84) (32,86)};
\addlegendentry{\texttt{LDS}}
\addplot[red!80!black, mark=triangle*, mark size=1.4pt, line width=.6pt] coordinates {(1,39) (2,41) (3,43) (4,45) (5,47) (6,49) (7,51) (8,53) (9,55) (10,57) (11,59) (12,61) (13,63) (14,65) (15,67) (16,69) (17,71) (18,73) (19,75) (20,77) (21,79) (22,81) (23,83) (24,85) (25,87) (26,89) (27,91) (28,93) (29,95) (30,97) (31,99) (32,101)};
\addlegendentry{\texttt{LDG}}
\end{axis}
\begin{axis}[at={(2.7cm,0)}, scale only axis, width=2.1cm, height=2.3cm,
  xmin=0, xmax=33, ymin=15, ymax=120, xtick={1,8,16,24,32}, minor xtick={1,...,32},
  ytick={20,40,...,120}, minor ytick={20,25,...,120},
  major tick length=3pt, minor tick length=1.4pt, tick align=outside, tick pos=left,
  tick label style={font=\scriptsize}, label style={font=\scriptsize},
  xlabel={$\mathit{bf}$}, xlabel style={yshift=1ex},
  title={(b) V100}, title style={font=\scriptsize, yshift=-1ex},
  yticklabels={},
]
\addplot[blue!80!black, mark=square*, mark size=1.2pt, line width=.6pt] coordinates {(1,20) (2,22) (3,24) (4,26) (5,28) (6,30) (7,32) (8,34) (9,36) (10,38) (11,40) (12,42) (13,44) (14,46) (15,48) (16,50) (17,52) (18,54) (19,56) (20,58) (21,60) (22,62) (23,64) (24,66) (25,68) (26,70) (27,72) (28,74) (29,76) (30,78) (31,80) (32,82)};
\addplot[red!80!black, mark=triangle*, mark size=1.4pt, line width=.6pt] coordinates {(1,37) (2,45) (3,50) (4,61) (5,54) (6,56) (7,57) (8,61) (9,60) (10,57) (11,58) (12,61) (13,60) (14,60) (15,59) (16,61) (17,61) (18,61) (19,61) (20,61) (21,60) (22,60) (23,60) (24,61) (25,61) (26,61) (27,61) (28,61) (29,60) (30,60) (31,60) (32,61)};
\end{axis}
\begin{axis}[at={(5.4cm,0)}, scale only axis, width=2.1cm, height=2.3cm,
  xmin=0, xmax=33, ymin=15, ymax=120, xtick={1,8,16,24,32}, minor xtick={1,...,32},
  ytick={20,40,...,120}, minor ytick={20,25,...,120},
  major tick length=3pt, minor tick length=1.4pt, tick align=outside, tick pos=left,
  tick label style={font=\scriptsize}, label style={font=\scriptsize},
  xlabel={$\mathit{bf}$}, xlabel style={yshift=1ex},
  title={(c) H100}, title style={font=\scriptsize, yshift=-1ex},
  yticklabels={},
]
\addplot[blue!80!black, mark=square*, mark size=1.2pt, line width=.6pt] coordinates {(1,24) (2,26) (3,28) (4,30) (5,32) (6,34) (7,36) (8,38) (9,40) (10,42) (11,44) (12,46) (13,48) (14,50) (15,52) (16,54) (17,56) (18,58) (19,60) (20,62) (21,64) (22,66) (23,68) (24,70) (25,72) (26,74) (27,76) (28,78) (29,80) (30,82) (31,84) (32,86)};
\addplot[red!80!black, mark=triangle*, mark size=1.4pt, line width=.6pt] coordinates {(1,39) (2,41) (3,43) (4,45) (5,47) (6,49) (7,51) (8,53) (9,55) (10,57) (11,59) (12,61) (13,63) (14,65) (15,67) (16,69) (17,71) (18,73) (19,75) (20,77) (21,79) (22,81) (23,83) (24,85) (25,87) (26,89) (27,91) (28,93) (29,95) (30,97) (31,99) (32,101)};
\end{axis}
\end{tikzpicture}}%
\caption{Latency of a single 4-byte load as its bank conflict grows on (a) the
A100, (b) the V100, and (c) the H100.}
\label{fig:bank}
\end{figure}
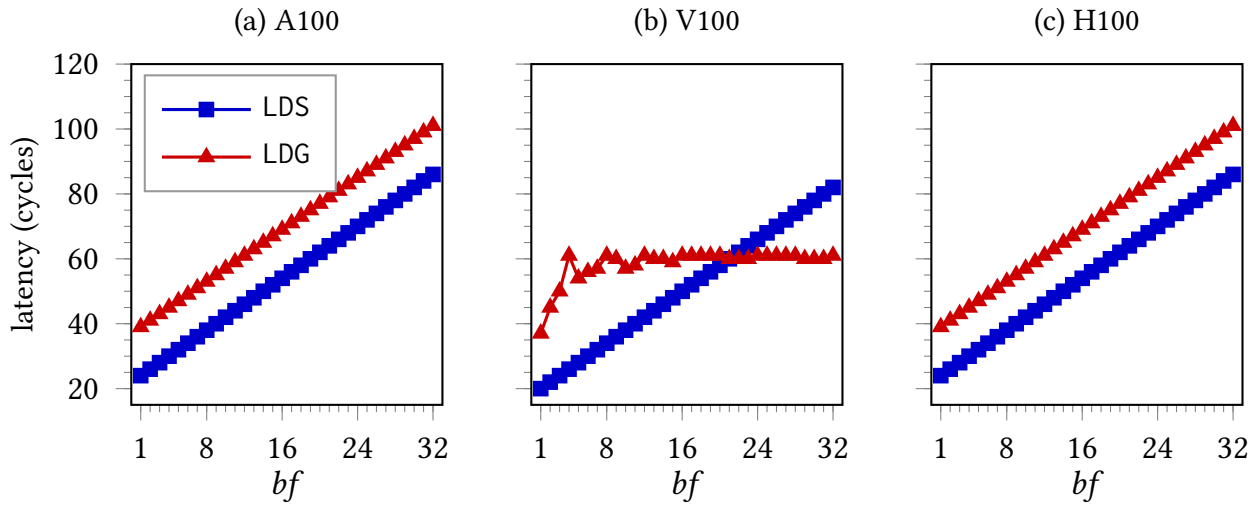

Almost every memory instruction of the SM accesses the banks
(\S\ref{sec:arch}). Prior work estimated about two cycles per additional way of bank conflict for shared loads only, on GPUs up to the A100~\cite{sun22,jiang19}; whether global loads pay it too has not
been measured. We study how the LSU resolves a conflict, whether each additional way of bank conflict costs exactly two cycles for a shared load, whether
global loads pay the same, and whether the cost is the same on the V100, the A100, and the H100.

We use the snippet of Fig.~\ref{fig:snippets}(a) once with an \texttt{LDS} instruction and once with an \texttt{LDG} instruction, raise $\mathit{bf}$ from 1 to 32,
and run it on the three GPUs (Fig.~\ref{fig:bank}). On the A100 and the H100, both
latencies grow by exactly two cycles per additional way of bank conflict, with identical readings on the two GPUs. However,
on the V100 the same \texttt{LDG} instruction grows only up to $\mathit{bf}=4$ and then
stays near 60 cycles, whereas the \texttt{LDS} instruction still pays two cycles per additional way of bank conflict.

\begin{finding}
\label{f:bank}
One bank access takes exactly two cycles, except for global loads on the V100. Since a bank conflict
is resolved by serialization and its cost is set by the busiest bank, a conflict of $\mathit{bf}$ ways adds an extra $2(\mathit{bf}-1)$ cycles.
\end{finding}

\subsection{Four Tag-Lookup Units Selected by an Address Hash}
\label{sec:tag}

\begin{figure}[t]
\centering
\includegraphics[width=\columnwidth]{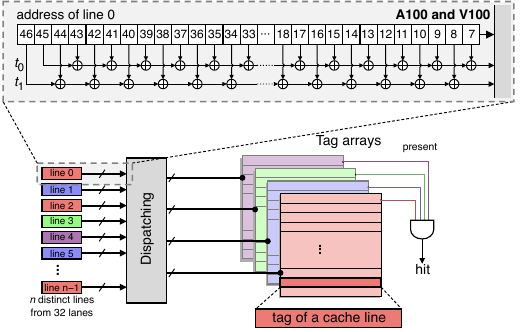}\\[1pt]
{\footnotesize (a)}\\[5pt]
\includegraphics[scale=1.15]{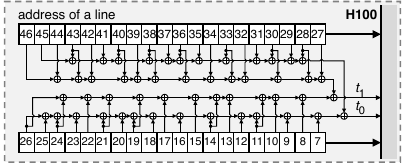}\\[1pt]
{\footnotesize (b)}%
\caption{Our model of the tag-lookup units, assuming that an earlier unit has
merged the addresses of the 32 lanes into $n$ distinct lines. (a) The GPU-independent model and the hash function of the A100 and V100. (b) The hash function of the H100. Colors indicate different hash results.}
\label{fig:tag}
\end{figure}

Before a global load can read the banks, the L1 cache must look up the tag of every cache line the load touches to determine whether the data it wants is in the L1 cache. NVIDIA states that this tag stage resolves at most four distinct lines per cycle~\cite{nvforum25tag}. If any four lines could be looked up together, a load would cost one cycle per four distinct lines, and the cost would depend only on how many lines it touches. The statement does not say so, and NVIDIA's patents suggest otherwise. A 2019 patent describes an L1 cache that splits the tag lookup over several tag banks, four in its example, so that a request can look up in one cycle only lines that fall in different tag banks~\cite{nvpat11080051}.

An earlier patent, filed in 2012, the year of
NVIDIA's Kepler GPUs, describes the tag unit of the texture pipeline, which on Kepler also served
read-only global loads through a texture cache that sat beside the L1 cache and the shared
memory~\cite{keplerwp}. However, since the V100, the L1 cache and the shared memory
are one on-chip SRAM that also serves the texture pipeline~\cite{v100wp}. That
patent spreads cache lines over four tag memories and picks a memory from two bits:
the XOR of all even-indexed address bits and the XOR of all odd-indexed address
bits~\cite{nvpat9348762}.
Whether the A100 assigns lines to tag-lookup units this way, and what a conflict
costs, has not been measured.

To find out, we use the snippet of Fig.~\ref{fig:snippets}(a) with a single
\texttt{LDG}, read the latency of the load directly, and choose the cache
line that each of its 32 lanes reads. All lanes hit in the L1 cache and no two of
them conflict on a bank ($\mathit{bf}=1$), so neither a miss nor a bank
conflict enters the reading. When the lanes read $L$ consecutive lines, the
latency rises from 32 at $L=1$, the cost of a single line without bank
conflicts, by one cycle every four lines to 39 at $L=32$, just as four lines
per cycle predicts.

However, when we keep 32 distinct lines and change only
where they lie in the address space, the latency varies from 39, the value for
32 consecutive lines, up to 63. The lines must therefore be assigned to the units by their addresses. The patent's two-bit XOR is one such assignment, and
the set-index functions of the L2 and L3 translation lookaside buffers of NVIDIA
GPUs~\cite{tunnels23} and of the Fermi L1 data cache~\cite{nugteren14} are also
XORs of address bits. We therefore assume that a line's unit is selected by XORs of address bits, and recover which
bits each XOR combines as follows. Two lines fall in the same unit exactly when these XORs take
the same values on both addresses, so every pair of colliding lines gives one linear equation
over the address bits; from enough such pairs we solve for the XORs, as was done for the slice
hash of Intel's last-level cache~\cite{irazoqui15} and for DRAM address mappings~\cite{knockknock25}. Jain
et al.\ also collected colliding pairs on GPUs, but searched over all candidate XORs~\cite{jain19}; we instead solve the equations by Gaussian elimination over GF(2), without a search.

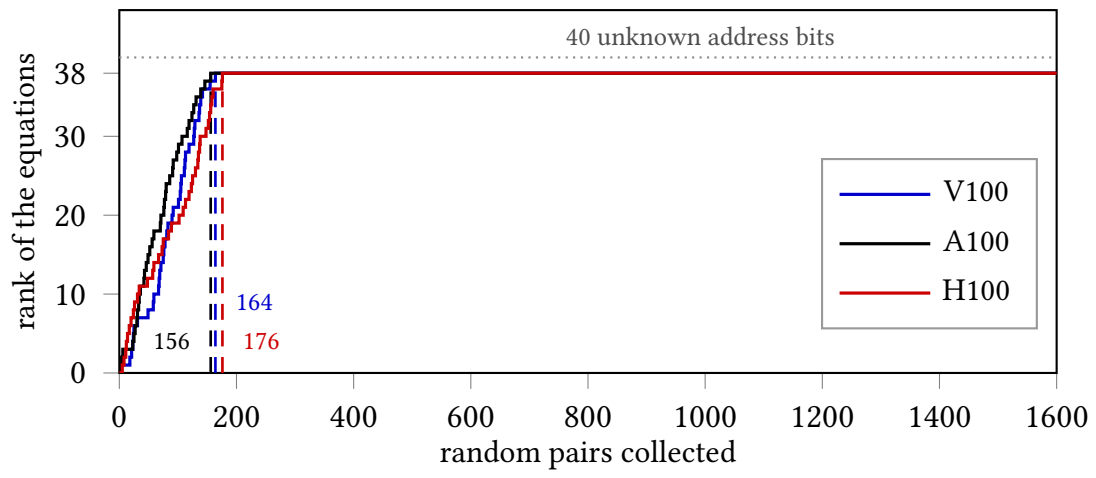
\begin{figure}[t]
\centering
\begin{tikzpicture}
\begin{axis}[scale only axis, width=6.2cm, height=2.4cm,
  xmin=0, xmax=1600, ymin=0, ymax=46, xtick={0,200,...,1600}, ytick={0,10,20,30,38},
  x tick label style={/pgf/number format/set thousands separator={}},
  major tick length=3pt, tick align=outside, tick pos=left,
  tick label style={font=\scriptsize}, label style={font=\scriptsize},
  xlabel={random pairs collected}, xlabel style={yshift=1ex},
  ylabel={rank of the equations}, ylabel style={yshift=-1ex},
  legend style={font=\scriptsize, draw=black!40, fill=white, at={(0.98,0.12)}, anchor=south east},
  v/.style={blue!80!black, line width=.6pt}, a/.style={black, line width=.6pt}, h/.style={red!80!black, line width=.6pt},
]
\draw[black!45, densely dotted] (axis cs:0,40) -- (axis cs:1600,40) node[pos=0.62, above=-1pt, font=\tiny, black!70] {40 unknown address bits};
\addplot[v, const plot] coordinates {(0,0) (4,1) (18,2) (21,3) (23,4) (24,5) (25,6) (31,7) (49,8) (58,9) (59,10) (67,11) (68,12) (69,13) (71,14) (75,15) (76,16) (78,17) (81,18) (83,19) (90,20) (92,21) (101,22) (104,23) (105,24) (106,25) (111,26) (112,27) (113,28) (119,29) (127,30) (128,31) (129,32) (136,33) (137,34) (139,35) (142,36) (155,37) (164,38) (1600,38)};
\addlegendentry{V100}
\draw[blue!80!black, densely dashed, line width=.5pt] (axis cs:164,0) -- (axis cs:164,38);
\node[blue!80!black, font=\tiny, anchor=west] at (axis cs:170,9) {164};
\addplot[a, const plot] coordinates {(0,0) (1,1) (4,2) (6,3) (23,4) (25,5) (27,6) (30,7) (31,8) (33,9) (34,10) (36,11) (42,12) (43,13) (46,14) (49,15) (52,16) (56,17) (59,18) (70,19) (71,20) (76,21) (77,22) (79,23) (80,24) (86,25) (91,26) (92,27) (98,28) (101,29) (107,30) (116,31) (119,32) (124,33) (127,34) (131,35) (139,36) (146,37) (156,38) (1600,38)};
\addlegendentry{A100}
\draw[black, densely dashed, line width=.5pt] (axis cs:156,0) -- (axis cs:156,38);
\node[black, font=\tiny, anchor=east] at (axis cs:150,4) {156};
\addplot[h, const plot] coordinates {(0,0) (5,1) (8,2) (11,3) (12,4) (14,5) (17,6) (20,7) (24,8) (26,9) (31,10) (34,11) (48,12) (57,13) (59,14) (67,15) (73,16) (75,17) (84,18) (89,19) (102,20) (109,21) (113,22) (119,23) (123,24) (125,25) (130,26) (134,27) (135,28) (137,29) (138,30) (148,31) (152,32) (154,33) (156,34) (158,35) (160,36) (175,37) (176,38) (1600,38)};
\addlegendentry{H100}
\draw[red!80!black, densely dashed, line width=.5pt] (axis cs:176,0) -- (axis cs:176,38);
\node[red!80!black, font=\tiny, anchor=west] at (axis cs:182,4) {176};
\end{axis}
\end{tikzpicture}%
\caption{Rank of the collision equations as random pairs are collected on the
V100, the A100, and the H100. Dashed lines mark the pair at which the rank
first reaches 38, which fixes the XORs; it then stays at 38 for all remaining
pairs.}
\label{fig:rank}
\end{figure}

We collect the colliding pairs with a plain CUDA kernel, without editing SASS
(Fig.~\ref{fig:cudasnip}). Thirty-one lanes read words of one cache line and
the remaining lane reads a word of another, all on distinct banks; both lines
are chosen at random, afresh for every measurement. Each lane reads its word
again and again, each load waiting for the value returned by the previous one,
so across 1{,}600 random pairs on the A100, the time of one load, averaged over
8{,}192 loads, takes only two values, 38.89 and 39.89 cycles: exactly one cycle more when the two lines share a unit. Two lines $x$
and $y$ share a unit exactly when, among the address bits used by each XOR, an
even number take different values in $x$ and $y$. Every collision therefore
gives one linear equation over $\mathrm{GF}(2)$ (the two-element field
$\{0, 1\}$, where addition is XOR) for each XOR, and
every XOR of the hash lies in the null space of the collected equations.
On all three GPUs, the rank of the collected equations reaches 38 after about
160 pairs and never grows over the remaining pairs (Fig.~\ref{fig:rank}).
With the 40 address bits $a_7$--$a_{46}$ and 38 linearly independent equations,
the XORs that satisfy every collision span a two-dimensional space; any two
linearly independent XORs in it split the lines into the same four groups.
Moreover, about one in four of the 1{,}600 pairs collides, as expected when two
XORs split the lines evenly over four units. Hence, we conclude that there are
four tag-lookup units.

Solving the equations, we find that the V100 and the A100 assign lines to their
four tag-lookup units by the same two XORs, those of the 2012 patent (top of Fig.~\ref{fig:tag}(a)): $t_0$ is
the XOR of all odd address bits and $t_1$ the XOR of all even ones. The H100,
however, uses two different XORs, which follow no simple pattern
(Fig.~\ref{fig:tag}(b)).

We summarize the model in Fig.~\ref{fig:tag}(a). Dispatching sends each distinct
line, according to its hash, to its tag-lookup unit. As in a typical cache, the tag-lookup
unit takes some of the address bits as the tag and checks whether it is present,
although we cannot determine which bits; the load hits only if every tag is
present.

Our tests only show which lines share a tag-lookup unit, not which unit it is. They
therefore cannot tell whether the hardware computes $t_0$ and $t_1$, or replaces
one of them with $t_0 \oplus t_1$. Such a replacement only renames the four tag
units: lines that share a unit before it still share one after it. Since the
cost depends only on which lines share a unit, the choice changes no prediction.

\begin{figure}[t]
\centering\scriptsize
\begin{tcolorbox}[enhanced, width=\columnwidth, nobeforeafter, colback=white, colframe=black, boxrule=0.6pt, arc=0pt, left=3pt, right=3pt, top=1pt, bottom=1pt]
\begin{Verbatim}[commandchars=\\\{\}]
\textcolor{codecmt}{// lane l < 31 points to word l of line x,}
\textcolor{codecmt}{// lane 31 points to word 31 of line y;}
\textcolor{codecmt}{// p[0] == 0 for every lane}
int v = 0;
for (int i = 0; i < 256; i++)  v = p[v]; \textcolor{codecmt}{// warm-up}
long long t0 = clock64();
for (int i = 0; i < 8192; i++) v = p[v]; \textcolor{codecmt}{// timed}
double r = (clock64() - t0) / 8192.0;
\end{Verbatim}
\end{tcolorbox}
\caption{CUDA code of the collision test.}
\label{fig:cudasnip}
\end{figure}

\begin{finding}
\label{f:tag}
When the L1 cache looks up the tags of a load, each distinct cache line goes to one
of four tag-lookup units chosen by two XORs of its address bits.
\end{finding}

\subsection{The Cost of the Tag Lookup}
\label{sec:tagcost}

With the hash known, we are able to determine how the tag-lookup units work
and what they cost: we pick addresses whose hashes send each line to the unit we
want, without bank conflicts. We
first put all $n$ distinct lines of a load into one unit, for $n$ from 1 to 32,
and time the load with the snippet of Fig.~\ref{fig:snippets}(a). On the A100
and the H100, the latency is $31+n$ cycles; on the V100, it is 27 cycles for one
line and $27+n$ cycles from two lines on. Each additional line in one unit thus
adds one cycle: a unit resolves one line per cycle. On the V100, the second line
adds two cycles and every further line one, for a reason we could not determine.

We then keep 32 distinct lines and change only how they are spread over the four units.
On the A100, the latency is 31 plus the number of lines in the busiest unit; how the other lines
are spread does not matter. This explains the range of \S\ref{sec:tag}: 32 lines spread evenly
over the units put eight lines in each and take eight cycles, 39 in total, whereas 32 lines in one
unit take 32 cycles, 63 in total. The V100 and the H100 follow the same rule in the layouts we
measured. The four units
therefore work in parallel, and the busiest one sets the cost of the tag lookup.

\begin{finding}
\label{f:tagcost}
Each tag-lookup unit resolves one line per cycle, so the tag-lookup cost of a load is the number of its lines that fall into the busiest unit.
\end{finding}

\subsection{The Latency of a Load}
\label{sec:formula}

The previous sections gave the two costs that depend on the addresses of a
load. In this section, we therefore explore the full latency of a load instruction.
Every experiment uses the snippet of Fig.~\ref{fig:snippets}(a) and varies one
or more of the memory space, the access width, and the address layout.

The experiments of \S\ref{sec:bank} show that on all three generations, the latency
of a 4-byte shared load is $\mathit{latency} = \mathit{base} + 2(\mathit{bf}-1)$,
where the base is the latency of the same load when its 32 lanes read consecutive
addresses (Table~\ref{tab:base}), and the $-1$ accounts for the one bank access
that the base already includes.

\begin{table}[t]
\caption{Base: the latency of a load whose 32 lanes read
consecutive addresses, in cycles. Global loads on the V100 are not covered.}
\label{tab:base}
\centering
\footnotesize
\begin{tabular}{l|ccc|ccc}
 & \multicolumn{3}{c|}{global} & \multicolumn{3}{c}{shared} \\
GPU & 4\,B & 8\,B & 16\,B & 4\,B & 8\,B & 16\,B \\ \hline
A100 & 32 & 36 & 44 & 24 & 28 & 36 \\
H100 & 32 & 35 & 41 & 24 & 27 & 33 \\
V100 & -- & -- & -- & 20 & 25 & 33 \\
\end{tabular}
\end{table}

To check whether the tag-lookup cost and the bank-access cost add for a global
load, we vary both in the same load. Since the tag-lookup cost is set by the
busiest unit (Finding~\ref{f:tagcost}), we put all its lines into one unit and raise their number from 1 to 32; we write $T$ for the number of lines in the busiest unit. For each $T$, we let $\mathit{bf}$ lanes read the same word of different lines, for
$\mathit{bf}$ from 1 to $T$, since a $\mathit{bf}$-way conflict needs
$\mathit{bf}$ lines. On the A100 and the H100, the latency is
\begin{equation}
\mathit{latency} = \mathit{base} + (T-1) + 2(\mathit{bf}-1)
\label{eq:latency}
\end{equation}
for every $T$ and $\mathit{bf}$. Global loads on the V100 are not covered,
since their bank conflicts do not cost two cycles per additional way of bank conflict (Finding~\ref{f:bank}).

An NVIDIA engineer states that the hardware splits a 16-byte shared store into
four groups of eight consecutive lanes and serves them one group at a
time~\cite{nvforum26}. To check this for loads, we give an 8-byte
load on the A100 a 16-way conflict among lanes 0--15 and another among lanes 16--31, in
different banks. In this experiment, the two conflicts cost 60 extra cycles, the sum of their separate costs, whereas
in a 4-byte load, the same two conflicts cost only 30 extra cycles in total.
This means that the LSU splits an 8-byte load into two groups of lanes, lanes 0--15 and
lanes 16--31, and serves them one after the other. When we vary the conflict
of each group, separately and together, up to 16 ways per group for 8-byte loads
and up to 8 ways per group for 16-byte loads, in both memory spaces, the cost is
always the sum over the groups, $2\sum_g(\mathit{bf}_g-1)$ cycles. For global
loads, we also put 1 to 32 distinct lines into one tag-lookup unit in each of
these settings: the conflict cost stays the same sum, and the latency grows by
one cycle per line, as for a 4-byte load (\S\ref{sec:tagcost}).

\begin{finding}
\label{f:formula}
The latency of a global load that hits in the L1 cache is the sum of a base, its tag-lookup
cost, and its bank-access cost. The latency of a shared load is the sum of a base and its
bank-access cost. The base is the latency of the same load with consecutive addresses.
\end{finding}

\subsection{Two Concurrent Access Channels}
\label{sec:chan}

\begin{figure}[t]
\centering
\input{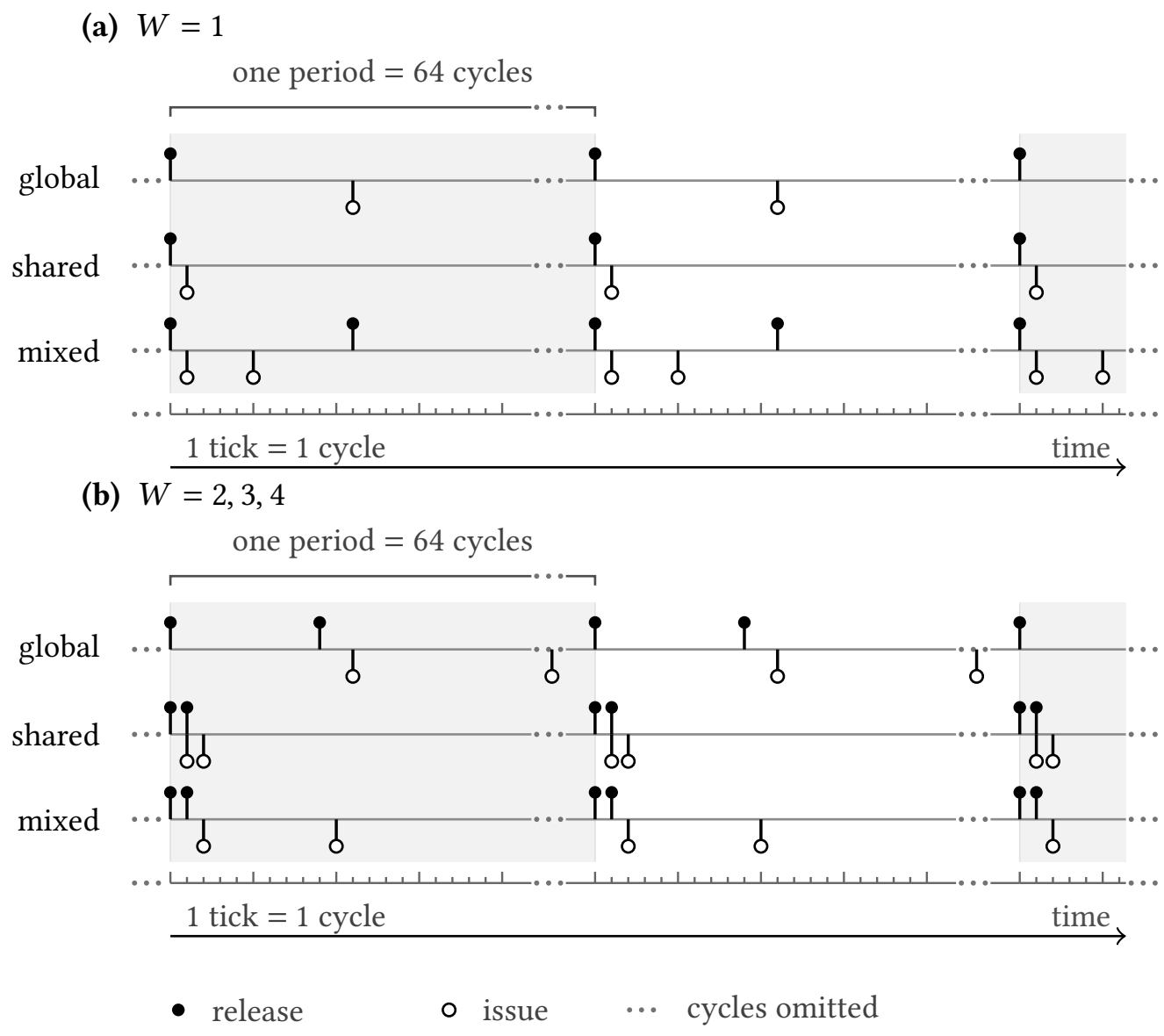}%
\caption{Release ($\bullet$) and issue ($\circ$) times of saturated sequences of
4-byte loads with 32-way bank conflicts on the H100. Each shaded or white band is
one period of 64 cycles. (a) A single warp issues and releases one load per period
with a pure global or pure shared sequence, and two with a mixed sequence. (b)
With two to four warps on different sub-cores, the SM issues and releases two
loads per period with every sequence.}
\label{fig:period}
\end{figure}

\S\ref{sec:bank} showed that one bank access takes two cycles. This latency
limits throughput most when the banks serve one load at a time. In that case, an
SM could complete just one access every two cycles. We therefore study whether the
banks can serve several loads at once, and if so, which ones.

To find out, we build a timeline (\S\ref{sec:cycletable}): we record the issue and
release cycles of every load in a sequence, and keep only the loads that issue after the LSU pipeline is saturated. We read these
cycles with the snippet of Fig.~\ref{fig:snippets}(b), lengthening the sequence one load
at a time: a global sequence consists only of \texttt{LDG}, a shared sequence only of
\texttt{LDS}, and a mixed sequence alternates between the two. Every load has a 32-way bank conflict, its 32 lanes all reading bank 0 in different lines; for a global load, its 32 cache lines are consecutive and therefore spread
evenly over the four tag-lookup units, so the pace is set by the slowest stage, the bank
access. We run each sequence with one to four warps on the A100 and H100. Fig.~\ref{fig:period} shows these timelines on H100, drawn as in Fig.~\ref{fig:snippets}(c). The figure is drawn at
$\mathit{bf}=32$, where the timeline repeats with a period of 64 cycles. For every $\mathit{bf}$ from 8 to 32, the
pattern repeats exactly every $2\mathit{bf}$ cycles, with the same numbers of
issues and releases per period. One period is therefore the time the banks
spend on one load: 32 accesses, two cycles each. To see how many loads the banks
serve at once, we count the releases ($\bullet$) in one period.

In most cases, two loads issue and two release per period, and this count is
the same in every period from the moment the buffer fills until the sequence of
loads ends. The only exception is a single warp that sends all its loads to one
memory space: it issues and releases only one load per period
(Fig.~\ref{fig:period}(a), global and shared rows). No row shows more than two.
A third or fourth warp changes nothing: the SM still releases two loads per
period for $W=2$, 3, and 4 (Fig.~\ref{fig:period}(b)).

Since each load occupies the banks for a whole period, two releases per period
mean two loads served at once. The SM thus serves at most two loads at a time, even though all of them use the same bank. Since one bank access takes two cycles (\S\ref{sec:bank}), each of the two services completes one access every two cycles, and the SM performs one bank access per cycle in total. We call these two services the concurrent access channels. Each channel can carry either global or shared loads. Two
warps use both channels when all loads are global, and also when all loads are
shared (Fig.~\ref{fig:period}(b)). For a single warp, however, the second channel is free, yet its loads to one memory space are served
one at a time; the same warp fills both channels once it alternates global and
shared loads (Fig.~\ref{fig:period}(a), mixed row).

\begin{finding}
\label{f:chan}
An SM has two concurrent access channels, but two loads are
served at the same time only if they come from different warps or access
different memory spaces. The two loads need not use different banks. Each channel completes one bank access every two cycles, so the SM performs at most one bank access per cycle.
\end{finding}

\subsection{One Load at a Time in the Tag-Lookup Units}
\label{sec:tagserial}

The four tag-lookup units work in parallel within one load
(\S\ref{sec:tagcost}). We now explore whether they can also serve different loads at the same time.
We use the same method as in \S\ref{sec:chan}, except that the loads have no bank
conflicts and the 32 lines of every load fall into one tag-lookup unit: with one
warp, four consecutive loads fall into the four different units, and with several
warps, the loads of each warp fall into a different unit.
If the units served different loads independently, each unit in use would release
one load every 32 cycles, so the SM would release up to four loads every 32 cycles. On the V100, the A100, and the H100, it releases
one load every 32 cycles for every $W$ from 1 to 4, so each warp waits $32W$ cycles between its
loads. When the 32 lines of each load are spread evenly over the units instead,
the SM releases one load every 8 cycles, again for every $W$: the interval is the
tag-lookup cost of one load (Finding~\ref{f:tagcost}).

\begin{finding}
\label{f:tagserial}
The tag-lookup units serve one load at a time for the whole SM.
\end{finding}

\subsection{When a Warp Starves}
\label{sec:buffer}

\begin{figure}[t]
\centering
\tikzset{no/.style={black, mark=*, mark size=1.0pt, line width=.5pt}, hi/.style={blue!80!black, mark=square*, mark size=0.95pt, line width=.5pt}, fi/.style={red!80!black, mark=triangle*, mark size=1.0pt, line width=.5pt}}% styles of the shared legend of (a)(b)
\begin{minipage}[t]{\columnwidth}\centering\ref*{gaplegA}\end{minipage}\\[1pt]
\begin{minipage}[b]{0.5\columnwidth}\centering\tikzset{no/.style={black, mark=*, mark size=1.0pt, line width=.5pt}, hi/.style={blue!80!black, mark=square*, mark size=0.95pt, line width=.5pt}, fi/.style={red!80!black, mark=triangle*, mark size=1.0pt, line width=.5pt}}
\begin{tikzpicture}[baseline=(current axis.south)]
\useasboundingbox (-1.15cm,-0.95cm) rectangle (3.08cm,2.35cm);
\begin{axis}[scale only axis, width=2.93cm, height=2.0cm, at={(0,0)}, anchor=south west, ymode=log, log basis y=2,
  xmin=3, xmax=33, ymin=20, ymax=52000,
  xtick={4,8,16,24,32}, minor xtick={4,8,...,32}, ytick={32,128,512,2048,8192,32768}, extra y ticks={64,256,1024,4096,16384}, extra y tick labels={}, extra y tick style={major tick length=1mm, grid=none},
  yticklabels={$2^5$,$2^7$,$2^9$,$2^{11}$,$2^{13}$,$2^{15}$}, tick align=outside, tick pos=left,
  tick label style={font=\fontsize{7}{8.5}\selectfont}, label style={font=\fontsize{7}{8.5}\selectfont},
  xlabel={$\mathit{bf}$}, ylabel={wait per \texttt{LDS}},
  ylabel style={yshift=-1ex}, xlabel style={yshift=1ex},
  grid=major, grid style={black!10, line width=.3pt},
  legend columns=3, legend cell align={left}, legend to name=gaplegA, legend style={font=\fontsize{7}{8}\selectfont, draw=none, fill=none, inner sep=0.5pt, nodes={inner sep=0.4pt}, /tikz/every odd column/.append style={column sep=0.5mm}, /tikz/every even column/.append style={column sep=2.5mm}},
  legend image code/.code={\draw[mark repeat=2, mark phase=2] plot coordinates {(0cm,0cm) (0.1cm,0cm) (0.2cm,0cm)};},
  no/.style={black, mark=*, mark size=1.0pt, line width=.5pt},
  hi/.style={blue!80!black, mark=square*, mark size=0.95pt, line width=.5pt},
  fi/.style={red!80!black, mark=triangle*, mark size=1.0pt, line width=.5pt},
]
\addplot[no] coordinates {(4,736) (5,984) (6,1234) (7,1481) (8,1730) (9,31477) (10,31277) (11,31327) (12,31477) (13,31516) (14,31362) (15,32299) (16,32432) (17,32499) (18,32445) (19,32455) (20,32424) (21,32486) (22,32385) (23,32388) (24,32486) (25,32455) (26,32284) (27,32300) (28,32366) (29,32318) (30,32380) (31,32379) (32,32330)};
\addlegendentry{no pause}
\addplot[hi] coordinates {(4,361) (5,487) (6,603) (7,720) (8,843) (9,1872) (10,2116) (11,2375) (12,2612) (13,2857) (14,3092) (15,3395) (16,3712) (17,3954) (18,4209) (19,4459) (20,4700) (21,4950) (22,5187) (23,5447) (24,5694) (25,5945) (26,6179) (27,6427) (28,6685) (29,6945) (30,7170) (31,7435) (32,7682)};
\addlegendentry{$p{+}6$}
\addplot[fi] coordinates {(4,48) (5,62) (6,75) (7,86) (8,94) (9,99) (10,113) (11,129) (12,139) (13,133) (14,161) (15,169) (16,178) (17,194) (18,191) (19,202) (20,226) (21,240) (22,241) (23,253) (24,265) (25,274) (26,286) (27,298) (28,310) (29,328) (30,332) (31,344) (32,355)};
\addlegendentry{inverse launch}
\end{axis}
\end{tikzpicture}\end{minipage}%
\begin{minipage}[b]{0.5\columnwidth}\centering\tikzset{no/.style={black, mark=*, mark size=1.0pt, line width=.5pt}, hi/.style={blue!80!black, mark=square*, mark size=0.95pt, line width=.5pt}, fi/.style={red!80!black, mark=triangle*, mark size=1.0pt, line width=.5pt}}
\begin{tikzpicture}[baseline=(current axis.south)]
\useasboundingbox (-1.15cm,-0.95cm) rectangle (3.08cm,2.35cm);
\begin{axis}[scale only axis, width=2.93cm, height=2.0cm, at={(0,0)}, anchor=south west, ymode=log, log basis y=2,
  xmin=3, xmax=33, ymin=20, ymax=52000,
  xtick={4,8,16,24,32}, minor xtick={4,8,...,32}, ytick={32,128,512,2048,8192,32768}, extra y ticks={64,256,1024,4096,16384}, extra y tick labels={}, extra y tick style={major tick length=1mm, grid=none},
  yticklabels={$2^5$,$2^7$,$2^9$,$2^{11}$,$2^{13}$,$2^{15}$}, tick align=outside, tick pos=left,
  tick label style={font=\fontsize{7}{8.5}\selectfont}, label style={font=\fontsize{7}{8.5}\selectfont},
  xlabel={$\mathit{bf}$}, ylabel={wait per \texttt{LDS}},
  ylabel style={yshift=-1ex}, xlabel style={yshift=1ex},
  grid=major, grid style={black!10, line width=.3pt},
  legend columns=3, legend cell align={left}, legend to name=gaplegB, legend style={font=\fontsize{7}{8}\selectfont, draw=none, fill=none, inner sep=0.5pt, nodes={inner sep=0.4pt}, /tikz/every odd column/.append style={column sep=0.5mm}, /tikz/every even column/.append style={column sep=2.5mm}},
  legend image code/.code={\draw[mark repeat=2, mark phase=2] plot coordinates {(0cm,0cm) (0.1cm,0cm) (0.2cm,0cm)};},
  no/.style={black, mark=*, mark size=1.0pt, line width=.5pt},
  hi/.style={blue!80!black, mark=square*, mark size=0.95pt, line width=.5pt},
  fi/.style={red!80!black, mark=triangle*, mark size=1.0pt, line width=.5pt},
]
\addplot[no] coordinates {(4,1003) (5,1258) (6,1514) (7,1771) (8,2025) (9,2281) (10,2535) (11,2791) (12,3048) (13,30884) (14,30944) (15,31350) (16,31339) (17,31386) (18,32437) (19,32378) (20,32620) (21,32472) (22,32575) (23,32395) (24,32297) (25,32536) (26,32330) (27,32572) (28,32542) (29,32479) (30,32527) (31,32493) (32,32354)};
\addlegendentry{no pause}
\addplot[hi] coordinates {(4,490) (5,618) (6,746) (7,875) (8,1000) (9,1129) (10,1256) (11,1384) (12,1510) (13,3116) (14,3356) (15,3627) (16,3855) (17,4100) (18,4571) (19,4820) (20,5081) (21,5337) (22,5597) (23,5850) (24,6088) (25,6343) (26,6613) (27,6862) (28,7119) (29,7378) (30,7632) (31,7884) (32,8142)};
\addlegendentry{$p{+}6$}
\addplot[fi] coordinates {(4,63) (5,70) (6,78) (7,84) (8,98) (9,109) (10,109) (11,117) (12,127) (13,136) (14,148) (15,155) (16,166) (17,172) (18,175) (19,181) (20,191) (21,201) (22,214) (23,210) (24,221) (25,231) (26,238) (27,247) (28,258) (29,270) (30,288) (31,300) (32,300)};
\addlegendentry{inverse launch}
\end{axis}
\end{tikzpicture}\end{minipage}\\[-2pt]
\begin{minipage}[t]{0.5\columnwidth}\centering\footnotesize (a)\end{minipage}%
\begin{minipage}[t]{0.5\columnwidth}\centering\footnotesize (b)\end{minipage}\\[2pt]
\begin{minipage}[b]{0.5\columnwidth}\centering\definecolor{hgrey}{HTML}{9E9E9E}\definecolor{hblue}{HTML}{6BAED6}
\begin{tikzpicture}[baseline=(current axis.south)]
\useasboundingbox (-1.15cm,-0.95cm) rectangle (3.08cm,2.35cm);
\begin{axis}[scale only axis, width=2.93cm, height=2.0cm, at={(0,0)}, anchor=south west, ybar, bar width=0.092cm, bar shift=0pt,
  xmin=-8192, xmax=8192, ymin=0, ymax=1700,
  xtick={-8192,-4096,0,4096,8192}, xticklabels={$-2^{13}$,,0,,$2^{13}$}, ytick={0,500,1000}, tick align=outside, tick pos=left,
  tick label style={font=\fontsize{7}{8.5}\selectfont}, label style={font=\fontsize{7}{8.5}\selectfont},
  xlabel={SM clock at issue, mod $2^{14}$}, ylabel={loads},
  ylabel style={yshift=-1ex}, xlabel style={yshift=1ex},
  axis on top, every axis plot/.append style={line width=.2pt},
  legend columns=1, legend cell align={left}, legend style={font=\fontsize{7}{8}\selectfont, draw=black!70, sharp corners, line width=.3pt, fill=white, inner sep=0.5pt, row sep=-1pt, nodes={inner sep=0.4pt}, at={(0.03,0.95)}, anchor=north west},
  legend image code/.code={\draw[#1] (0cm,-0.05cm) rectangle (0.2cm,0.05cm);},
]
\addplot[fill=hgrey, draw=black, line width=.2pt] coordinates {(-7936,20) (-7424,43) (-6912,28) (-6400,46) (-5888,27) (-5376,47) (-4864,36) (-4352,49) (-3840,31) (-3328,40) (-2816,40) (-2304,51) (-1792,32) (-1280,51) (-768,31) (-256,41) (256,29) (768,47) (1280,43) (1792,64) (2304,30) (2816,45) (3328,26) (3840,45) (4352,27) (4864,49) (5376,36) (5888,60) (6400,29) (6912,44) (7424,34) (7936,54)};
\addlegendentry{$\mathit{bf}=8$, 1275 loads}
\addplot[fill=hblue, draw=black, line width=.2pt] coordinates {(-7936,0) (-7424,0) (-6912,0) (-6400,0) (-5888,0) (-5376,0) (-4864,0) (-4352,0) (-3840,0) (-3328,0) (-2816,0) (-2304,0) (-1792,0) (-1280,0) (-768,0) (-256,0) (256,1230) (768,0) (1280,0) (1792,0) (2304,0) (2816,0) (3328,0) (3840,0) (4352,0) (4864,0) (5376,0) (5888,0) (6400,0) (6912,0) (7424,0) (7936,0)};
\addlegendentry{$\mathit{bf}=32$, 1230 loads}
\draw[black, dashed, line width=.4pt] (axis cs:0,0) -- (axis cs:0,1300);
\end{axis}
\end{tikzpicture}\end{minipage}%
\begin{minipage}[b]{0.5\columnwidth}\centering\definecolor{hgrey}{HTML}{9E9E9E}\definecolor{hblue}{HTML}{6BAED6}
\begin{tikzpicture}[baseline=(current axis.south)]
\useasboundingbox (-1.15cm,-0.95cm) rectangle (3.08cm,2.35cm);
\begin{axis}[scale only axis, width=2.93cm, height=2.0cm, at={(0,0)}, anchor=south west, ybar, bar width=0.092cm, bar shift=0pt,
  xmin=-8192, xmax=8192, ymin=0, ymax=1700,
  xtick={-8192,-4096,0,4096,8192}, xticklabels={$-2^{13}$,,0,,$2^{13}$}, ytick={0,500,1000}, tick align=outside, tick pos=left,
  tick label style={font=\fontsize{7}{8.5}\selectfont}, label style={font=\fontsize{7}{8.5}\selectfont},
  xlabel={SM clock at issue, mod $2^{14}$}, ylabel={loads},
  ylabel style={yshift=-1ex}, xlabel style={yshift=1ex},
  axis on top, every axis plot/.append style={line width=.2pt},
  legend columns=1, legend cell align={left}, legend style={font=\fontsize{7}{8}\selectfont, draw=black!70, sharp corners, line width=.3pt, fill=white, inner sep=0.5pt, row sep=-1pt, nodes={inner sep=0.4pt}, at={(0.03,0.95)}, anchor=north west},
  legend image code/.code={\draw[#1] (0cm,-0.05cm) rectangle (0.2cm,0.05cm);},
]
\addplot[fill=hgrey, draw=black, line width=.2pt] coordinates {(-7936,65) (-7424,25) (-6912,33) (-6400,62) (-5888,110) (-5376,54) (-4864,38) (-4352,1) (-3840,8) (-3328,5) (-2816,49) (-2304,9) (-1792,23) (-1280,28) (-768,52) (-256,94) (256,71) (768,21) (1280,38) (1792,62) (2304,97) (2816,54) (3328,21) (3840,0) (4352,20) (4864,12) (5376,26) (5888,5) (6400,22) (6912,31) (7424,50) (7936,89)};
\addlegendentry{$\mathit{bf}=8$, 1275 loads}
\addplot[fill=hblue, draw=black, line width=.2pt] coordinates {(-7936,0) (-7424,0) (-6912,0) (-6400,0) (-5888,0) (-5376,0) (-4864,0) (-4352,0) (-3840,0) (-3328,0) (-2816,0) (-2304,0) (-1792,0) (-1280,0) (-768,0) (-256,0) (256,1050) (768,75) (1280,0) (1792,0) (2304,0) (2816,0) (3328,0) (3840,0) (4352,0) (4864,0) (5376,0) (5888,0) (6400,0) (6912,0) (7424,0) (7936,0)};
\addlegendentry{$\mathit{bf}=32$, 1125 loads}
\draw[black, dashed, line width=.4pt] (axis cs:0,0) -- (axis cs:0,1300);
\end{axis}
\end{tikzpicture}\end{minipage}\\[-2pt]
\begin{minipage}[t]{0.5\columnwidth}\centering\footnotesize (c)\end{minipage}%
\begin{minipage}[t]{0.5\columnwidth}\centering\footnotesize (d)\end{minipage}
\caption{(a), (b) Average wait per \texttt{LDS} of the second kernel's warp that
shares its sub-core with the first kernel's warp, which is in a different block,
on (a) the A100 and (b) the H100. (c), (d) SM clock at which each load of that
warp issues, modulo $2^{14}$, when the warp spins a random time before
each load, with the first kernel's loads at
$\mathit{bf}=8$ (grey) and $\mathit{bf}=32$ (blue), on (c) the A100 and (d) the
H100.}
\label{fig:inflight_k}
\end{figure}
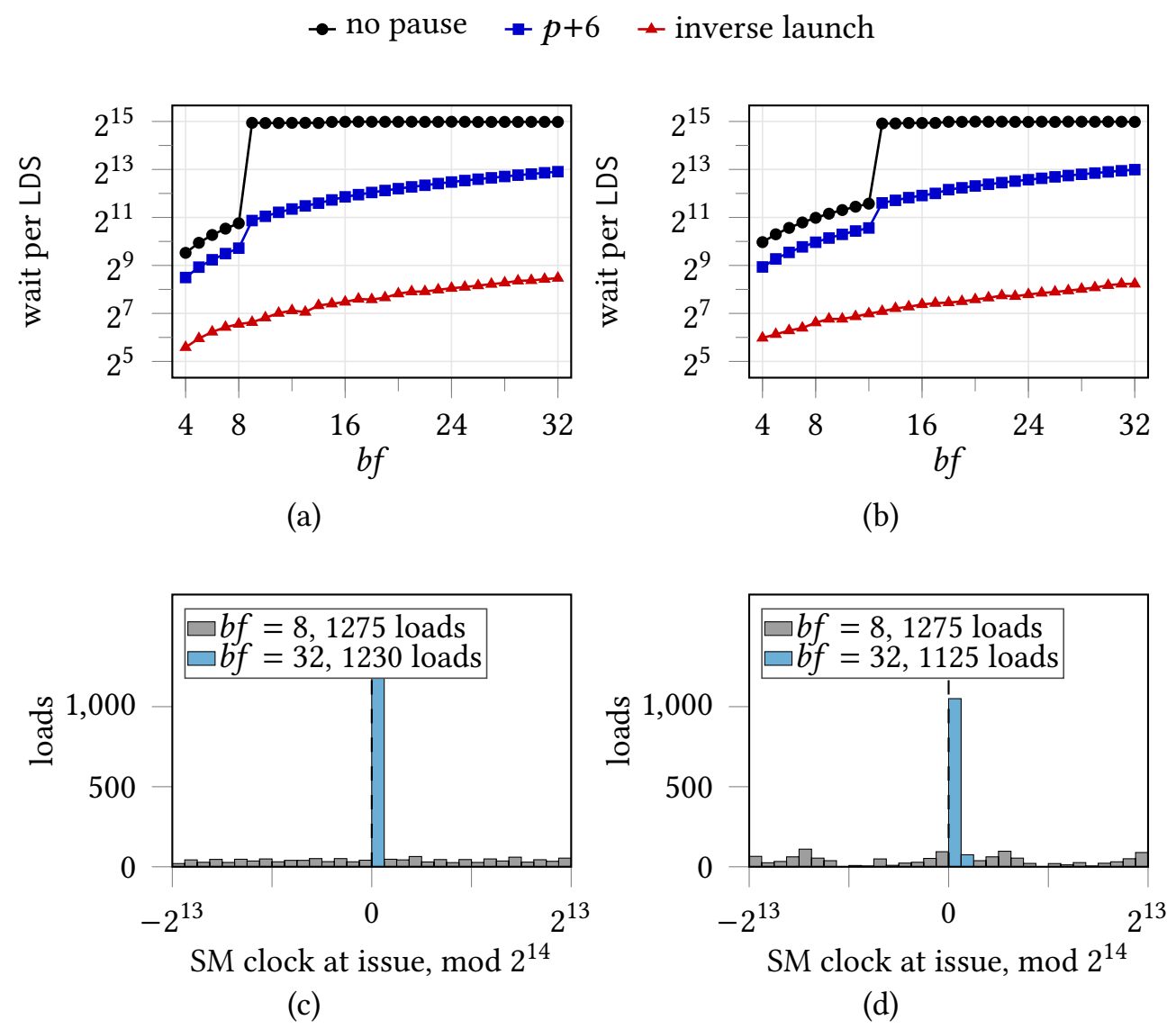

Knowing the cost of a bank conflict (\S\ref{sec:bank}) and the tag hash (\S\ref{sec:tag}), we
can choose the banks and the tag-lookup unit of a load instruction, and thus how many cycles it
occupies them. This lets us keep the LSU pipeline full. We finally explore whether a warp can starve while the
pipeline is full. We launch two kernels with one block per
SM each. In the first, a single warp keeps the LSU pipeline full with a loop of 128
consecutive \texttt{LDS} with $\mathit{bf}$-way bank conflicts, followed only
by the loop counter and the branch. The conflicts make each load cost
$p=2(\mathit{bf}-1)$ cycles on average (\S\ref{sec:chan}), so one loop takes
around $128p$ cycles when the loads issue back to back. The 128 loads are independent of one
another, and each has a stall count of one. In the second, one warp per sub-core
executes 32 \texttt{LDS}. We read \texttt{clock64()} before and after each load, and take the
difference as its timing. This
kernel is written in CUDA and compiled with \texttt{nvcc -O3} without editing, so
its control fields are those the compiler sets. Since the clock read after the load does
not use the loaded value, the compiler lets it wait on no DC, so the difference ends once
the load has issued, not when its data return: it is the time the load waited to
issue. Fig.~\ref{fig:inflight_k}(a) shows the average of this wait per load of
the warp that shares its sub-core with the first kernel's warp.

We first run the two kernels on the A100, launching the second after the first.
The black line in Fig.~\ref{fig:inflight_k}(a) shows the average wait per load of
the second kernel's warp. For $\mathit{bf}\le 8$, the wait is about one loop of
the first warp, $128p$ cycles: one load of the second warp enters per loop. For
$\mathit{bf}\ge 9$, it is about $2^{15}$ cycles. Within its loop, the first warp
always has a load ready, so a load waits only because the pipeline is full. The
only place where it has no instruction ready is the end of the loop, where we
find that the loop-control logic leaves it idle for about 17 cycles, a little
longer than the service time of one load at $\mathit{bf}=8$. To verify that an idle gap of this length lets a load in, we make the first warp pause in the middle of its loop for slightly longer than the
service time of one load, $p+6$ cycles (blue line in Fig.~\ref{fig:inflight_k}(a)): the second warp then enters twice per loop, at
the pause and at the loop end, so the wait is $64p$. Above the step the loop end
no longer lets it in and the pause is the only entry, so the wait is one loop.
Conversely, with a pause of $p-2$ cycles no load enters, and the wait is the same
as without a pause. The order of launch decides who waits:
when we launch the second kernel before the first, its warp on the shared
sub-core waits only a few hundred cycles per load, orders of magnitude less. The
other cases, namely warps of the same or of another block on the other three
sub-cores and warps of the same block on the same sub-core, show no comparably
long wait.

We also ran this test on the V100 and the H100. On the V100 we do not observe
the long wait. As Fig.~\ref{fig:inflight_k}(b) shows, the H100 behaves exactly as the
A100; since the loop end of this kernel leaves the first warp idle for about 25 cycles,
the step appears at $\mathit{bf}=13$ instead of 9.

\begin{finding}
When the LSU pipeline is full, a warp that always has a load ready keeps out the loads of the
warp of a block launched after its own on the same sub-core. A load of the later block gets
through only when that warp has no load ready for about the service time of one load.
\end{finding}

To explore where the $2^{15}$ comes from, we change the second kernel: it keeps one
warp per SM, on the first kernel's sub-core, which spins a random time before each
load, so that each load starts waiting at a random moment. We run the first kernel at
$\mathit{bf}=8$ and at $\mathit{bf}=32$, and collect the SM clock at which every load
of the second kernel issues, on every SM. Fig.~\ref{fig:inflight_k}(c), (d) show these
clocks modulo $2^{14}$, on the A100 and the H100, respectively (grey for
$\mathit{bf}=8$, blue for $\mathit{bf}=32$). At $\mathit{bf}=8$, as explained above, the loads issue at the end of the first
warp's loop, whose timing is set by when that warp started running and not by
the SM clock. This start differs from SM to SM, so pooled over all SMs the issue
instants fill the whole range. At $\mathit{bf}=32$, by contrast, we observe that
every load issues within a few tens to a few hundred cycles after a multiple of
$2^{14}$ of the SM clock, on every SM and on both GPUs, although the spin before
it is random. Since we record when each load begins to wait and when it issues, we also know how
long it waited. We find that no load issues at the first multiple after it
began to wait: the wait is almost never shorter than $2^{14}$ cycles and never longer
than about $2^{15}$. The issue of a waiting load is thus aligned to
the SM clock, not to the moment it began to wait: it issues at the first
multiple of $2^{14}$ of the clock at which it has already waited about $2^{14}$
cycles.

We can now explain the ceiling in Fig.~\ref{fig:inflight_k}(a) and (b). Above the
step, the loop end no longer lets a load in, so every load of the warp is kept out.
By the rule above, each of them issues right after a multiple of $2^{14}$. The next
load thus begins to wait right after that multiple. At the following multiple it has
waited less than $2^{14}$ cycles, so it issues at the one after, about $2^{15}$ cycles
after the previous load.

\begin{finding}
A load that is kept out issues no later than the first multiple of $2^{14}$ of
the SM clock after it has waited about $2^{14}$ cycles.
\end{finding}

\section{Implications}
\label{sec:impl}

In this section, we explore how the LSU components studied above slow a
colocated workload (\S\ref{sec:impl-dos}); form a covert channel across
processes (\S\ref{sec:impl-covert}); and expose timing side channels from
which we recover a full AES-128 key (\S\ref{sec:impl-sc}).

\subsection{Performance Degradation through Contention}
\label{sec:impl-dos}

In \S\ref{sec:buffer} we found that while the warp of an earlier-launched block
keeps the LSU pipeline full by issuing continuously, a warp of a block launched
after it on the same sub-core is nearly starved: it issues a load only about once
every $2^{15}$ cycles. A malicious co-tenant can turn this starvation
into a denial-of-service attack: by keeping the pipeline full it starves the loads
of any victim that shares its SM.

We assume a co-location threat model based on prior GPU performance
attacks~\cite{yandrofski22,wang25}, also common in covert
channels~\cite{naghib17,ahnnoc21}: the attacker and victim are two MPS processes
on the same GPU, co-resident on the same SMs and sharing no data; the attacker
runs the same back-to-back \texttt{LDS} loop as in \S\ref{sec:buffer}, but with 512 instead of
128 \texttt{LDS} per loop, and needs no privilege. Because we found that the order of launch decides who starves
(\S\ref{sec:buffer}), the attacker keeps its warp resident before the victim's.
In practice, typical kernels occupy as many SMs as possible for throughput, as NVIDIA
recommends~\cite{cudabp}, so one attacker warp per SM is a realistic
configuration. Furthermore, for stealth the attacker can launch as little as possible: when the
victim covers every SM, a single block of one warp suffices; otherwise it
launches a few more blocks so that at least one shares an SM with the victim.

We first launch our attack kernel with one warp per SM alongside an NVIDIA library
kernel; a cuBLAS kernel slows by up to $1221\times$ on the A100 and $563\times$
on the H100 (Table~\ref{tab:dos}), because each of its co-located loads now
incurs the enormous issue delay we explored in \S\ref{sec:buffer}. By contrast, on the V100 the slowdown is far smaller,
consistent with our finding that it does not starve (\S\ref{sec:buffer}).

To improve stealth, we shrink the attack to a single co-resident warp. Where the
victim leaves SMs idle, the attacker launches a few extra blocks so that one of them still
shares an SM with the victim. This single warp leaves a minimal footprint: its loads
touch only on-chip shared memory, so it moves no data through the L2 or DRAM. This footprint is wildly disproportionate to the damage:
one warp slows the \emph{whole} kernel, because a kernel ends only when its last
block finishes, so starving the single co-located block holds up the entire
kernel. Such a single warp, on the one SM a victim block occupies, slows
SGEMM ($n=512$) by $750\times$ on the A100 and $557\times$ on the H100 (Table~\ref{tab:dos}).

The attack fails in two cases: when the victim's own block monopolizes a shared
sub-core resource such as the register file, and when the victim makes little use
of the LSU, such as our FFMA loop, a hand-written kernel with no loads
at all (Table~\ref{tab:dos}). NVIDIA's Multi-Instance GPU (MIG) also stops the attack by giving each tenant its own SMs, at
the cost of fine-grained sharing. A targeted fix would instead make the issue
arbitration of memory instructions fair across warps on the same sub-core, whether from
different blocks or from different MPS processes, when an older warp always has a load ready
and the memory pipeline is full, so that it does not starve a younger warp.

\begin{implication}
Under MPS, an unprivileged co-tenant warp that shares a sub-core with the victim, starts before
it, and always has a load ready, can degrade the victim kernel by tens to hundreds of times.
\end{implication}

\begin{table}[t]
\caption{Slowdown of a colocated library kernel. Left value: one attacker warp on every SM;
right value: a single attacker warp on the SM of a victim block. Problem sizes: SGEMM $n=512$ and $2048$; SDOT
$2^{24}$ elements; FFT $2^{22}$ points (1-D) and $1024\times1024$ (2-D); SpMV on a banded matrix
($n=2^{20}$, 7 nonzeros per row) and on a matrix with random column indices ($n=2^{18}$, 16
nonzeros per row); sort $2^{22}$ keys; scan and reduce $2^{24}$ elements; gather $2^{22}$ elements.}
\label{tab:dos}
\centering
\footnotesize
\begin{tabular}{ll|c|c|c}
Library & Routine & H100 & A100 & V100 \\ \hline
cuBLAS   & SGEMM, $n=512$      & 557 / 557   & 1221 / 750  & 7.2 / 3.7 \\
cuBLAS   & SGEMM, $n=2048$     & 563 / 563   & 767 / 347   & 6.5 / 1.2 \\
cuBLAS   & SDOT                & 60.8 / 27.1 & 41.8 / 18.9 & 1.9 / 1.4 \\
cuFFT    & 1-D C2C             & 143 / 35.2  & 130 / 25.4  & 2.4 / 1.1 \\
cuFFT    & 2-D C2C             & 169 / 98.3  & 252 / 132   & 2.6 / 1.3 \\
cuSPARSE & SpMV, random        & 62.1 / 21.5 & 92.1 / 19.0 & 4.8 / 1.6 \\
cuSPARSE & SpMV, banded        & 94.8 / 23.4 & 144 / 17.7  & 6.0 / 1.6 \\
CUB      & RadixSort           & 229 / 87.6  & 440 / 85.1  & 13.6 / 2.0 \\
CUB      & InclusiveSum        & 150 / 25.3  & 206 / 10.9  & 12.7 / 2.2 \\
CUB      & Reduce              & 20.2 / 5.3  & 20.2 / 2.5  & 4.8 / 1.0 \\
Thrust   & gather              & 6.7 / 1.0   & 3.6 / 1.1   & 1.1 / 1.1 \\
ours     & FFMA loop, no loads & 1.4 / 1.4   & 1.2 / 1.0   & 1.2 / 1.0 \\
\end{tabular}
\end{table}

\subsection{A Cross-Process Covert Channel}
\label{sec:impl-covert}

In this section, we assume a co-location threat model that prior GPU covert channels have
repeatedly used~\cite{naghib17,ahnnoc21,nayak21,wang25,miao24}: two processes that are not
allowed to communicate are co-resident on the same GPU and signal through a shared
microarchitectural resource without sharing any data. We realize this model under MPS, as
in~\cite{nayak21,wang25,miao24}: our sender and receiver are two MPS processes co-resident
on the same SMs.

We develop two covert channels based on our findings, one on the two concurrent bank-access
channels (\S\ref{sec:chan}) and one on the four tag-lookup units (\S\ref{sec:tag}). Each is a shared, SM-level resource of the load instructions: for two processes or kernels
co-located on one SM, while one keeps the two bank-access channels busy, the other's loads
take longer to release; the tag-lookup unit behaves the same way.

\begin{algorithm}[t]
\DontPrintSemicolon
\SetKwProg{Proc}{Procedure}{}{}
\KwIn{SM index $i$;\ slot length $T$;\ message length $L$}
\Proc{$\mathrm{Sender}_i$}{
  $\mathrm{Synchronization}()$\;
  \While{\textnormal{true}}{
    $k \gets (\mathtt{globaltimer}() / T) \bmod L$\;
    \eIf{$M_i[k] = 1$}{
      $\mathrm{contend}()$\;
    }{
      do nothing\;
    }
  }
}
\Proc{$\mathrm{Receiver}_i$}{
  $\mathrm{Synchronization}()$\;
  $R_i[0{:}L] \gets 0$;\ \ $N_i[0{:}L] \gets 0$;\ \ $\mathit{covered} \gets 0$\;
  \While{$\mathit{covered} < L$}{
    $c_0 \gets \mathtt{clock}()$\;
    $\mathrm{probe}()$\;
    $lat \gets \mathtt{clock}() - c_0$\;
    \If{$\mathit{lane} = 0$}{
      $k \gets (\mathtt{globaltimer}() / T) \bmod L$\;
      \lIf{$N_i[k] = 0$}{$\mathit{covered} \gets \mathit{covered} + 1$}
      $R_i[k] \gets R_i[k] + lat$;\ \ $N_i[k] \gets N_i[k] + 1$\;
    }
  }
  \tcp{on the host}
  $\theta \gets \mathrm{median}_{k}\ R_i[k]/N_i[k]$\;
  \lFor{$k \gets 0$ \KwTo $L-1$}{$\widehat{M}_i[k] \gets 1\ \textnormal{if}\ R_i[k]/N_i[k] > \theta\ \textnormal{else}\ 0$}
}
\caption{Covert-channel protocol.}
\label{alg:covert}
\end{algorithm}

\begin{figure}[t]
\centering
\begin{tikzpicture}
\begin{axis}[hide axis, scale only axis, width=5cm, height=0.45cm,
  xmin=0, xmax=1, ymin=0, ymax=1,
  legend columns=3, legend style={font=\fontsize{6.5}{7.5}\selectfont, draw=black!70, sharp corners, line width=.3pt, fill=white, inner sep=1.4pt, /tikz/every even column/.append style={column sep=2.2mm}, at={(0.5,0.5)}, anchor=center},
  legend image code/.code={\draw[mark repeat=2, mark phase=2] plot coordinates {(0cm,0cm) (0.13cm,0cm) (0.26cm,0cm)};},
  v/.style={blue!80!black, mark=square*, mark size=1pt, line width=.6pt},
  a/.style={black, mark=*, mark size=1pt, line width=.6pt},
  h/.style={red!80!black, mark=triangle*, mark size=1.3pt, line width=.6pt},
]
\addlegendimage{v}\addlegendentry{V100}
\addlegendimage{a}\addlegendentry{A100}
\addlegendimage{h}\addlegendentry{H100}
\end{axis}
\end{tikzpicture}\\[2pt]
\begin{minipage}[b]{0.5\columnwidth}\centering\begin{tikzpicture}[baseline=(current axis.south)]
\begin{axis}[scale only axis, width=2.64cm, height=2.43cm,
  ymode=log, xmin=5, xmax=46, ymin=0.001, ymax=60,
  xtick={10,20,30,40}, minor xtick={5,10,...,45},
  ytick={0.01,0.1,1,10}, yticklabels={0.01,0.1,1,10},
  major tick length=3pt, minor tick length=1.6pt, tick align=outside,
  tick pos=left, tick label style={font=\scriptsize}, label style={font=\scriptsize},
  xlabel={bandwidth (Mb/s)}, ylabel={bit error rate (\%)},
  ylabel style={yshift=-1ex}, xlabel style={yshift=1ex},
  grid=major, grid style={black!10, line width=.3pt},
  v/.style={blue!80!black, mark=square*, mark size=0.88pt, line width=.6pt},
  a/.style={black, mark=*, mark size=0.88pt, line width=.6pt},
  h/.style={red!80!black, mark=triangle*, mark size=1.12pt, line width=.6pt},
]
\draw[dashed, black!45, line width=.3pt] (axis cs:5,1) -- (axis cs:46,1);
\addplot[v] coordinates {(8.000,0.0100) (13.333,1.4310) (16.000,2.2560) (17.778,4.0580) (20.000,7.3780) (22.857,16.2600) (26.667,24.5800)};
\addplot[a] coordinates {(10.800,0.0100) (18.000,0.0100) (21.600,1.7530) (24.000,6.9820) (27.000,4.7460) (30.857,18.0420) (36.000,27.5680)};
\addplot[h] coordinates {(13.200,0.0015) (22.000,0.0190) (26.400,0.0100) (29.333,0.2880) (33.000,0.8450) (37.714,10.6300) (44.000,7.3880)};
\end{axis}
\end{tikzpicture}\end{minipage}%
\begin{minipage}[b]{0.5\columnwidth}\centering\begin{tikzpicture}[baseline=(current axis.south)]
\begin{axis}[scale only axis, width=2.64cm, height=2.43cm,
  ymode=log, xmin=5, xmax=46, ymin=0.001, ymax=60,
  xtick={10,20,30,40}, minor xtick={5,10,...,45},
  ytick={0.01,0.1,1,10}, yticklabels={0.01,0.1,1,10},
  major tick length=3pt, minor tick length=1.6pt, tick align=outside,
  tick pos=left, tick label style={font=\scriptsize}, label style={font=\scriptsize},
  xlabel={bandwidth (Mb/s)}, ylabel={},
  ylabel style={yshift=-1ex}, xlabel style={yshift=1ex},
  grid=major, grid style={black!10, line width=.3pt},
  v/.style={blue!80!black, mark=square*, mark size=0.88pt, line width=.6pt},
  a/.style={black, mark=*, mark size=0.88pt, line width=.6pt},
  h/.style={red!80!black, mark=triangle*, mark size=1.12pt, line width=.6pt},
]
\draw[dashed, black!45, line width=.3pt] (axis cs:5,1) -- (axis cs:46,1);
\addplot[v] coordinates {(8.000,0.0100) (16.000,0.0290) (17.778,0.0490) (20.000,0.9330) (22.857,7.5190) (26.667,12.2170)};
\addplot[a] coordinates {(10.800,0.0190) (21.600,0.0100) (24.000,0.0100) (27.000,0.0980) (30.857,3.0570) (36.000,5.8010)};
\addplot[h] coordinates {(13.200,0.0100) (26.400,0.0190) (29.333,0.0290) (33.000,0.0100) (37.714,0.0490) (44.000,0.1850)};
\end{axis}
\end{tikzpicture}\end{minipage}\\[-2pt]
\begin{minipage}[t]{0.5\columnwidth}\centering\footnotesize (a) bank --- aggregate\end{minipage}%
\begin{minipage}[t]{0.5\columnwidth}\centering\footnotesize (b) tag --- aggregate\end{minipage}\\[4pt]
\begin{minipage}[b]{0.5\columnwidth}\centering\begin{tikzpicture}[baseline=(current axis.south)]
\begin{axis}[scale only axis, width=2.64cm, height=2.43cm,
  ymode=log, xmin=90, xmax=345, ymin=0.001, ymax=60,
  xtick={100,200,300}, minor xtick={100,150,...,350},
  ytick={0.01,0.1,1,10}, yticklabels={0.01,0.1,1,10},
  major tick length=3pt, minor tick length=1.6pt, tick align=outside,
  tick pos=left, tick label style={font=\scriptsize}, label style={font=\scriptsize},
  xlabel={per-SM rate (kb/s)}, ylabel={bit error rate (\%)},
  ylabel style={yshift=-1ex}, xlabel style={yshift=1ex},
  grid=major, grid style={black!10, line width=.3pt},
  v/.style={blue!80!black, mark=square*, mark size=0.88pt, line width=.6pt},
  a/.style={black, mark=*, mark size=0.88pt, line width=.6pt},
  h/.style={red!80!black, mark=triangle*, mark size=1.12pt, line width=.6pt},
]
\draw[dashed, black!45, line width=.3pt] (axis cs:90,1) -- (axis cs:345,1);
\addplot[v] coordinates {(100.000,0.0100) (111.111,0.0100) (125.000,0.0100) (142.857,0.0100) (166.667,1.4310) (200.000,2.2560) (222.222,4.0580) (250.000,7.3780) (285.714,16.2600) (333.333,24.5800)};
\addplot[a] coordinates {(100.000,0.0100) (111.111,0.0100) (125.000,0.0015) (142.857,0.0015) (166.667,0.0100) (200.000,1.7530) (222.222,6.9820) (250.000,4.7460) (285.714,18.0420) (333.333,27.5680)};
\addplot[h] coordinates {(100.000,0.0015) (111.111,0.0100) (125.000,0.0100) (142.857,0.0100) (166.667,0.0190) (200.000,0.0100) (222.222,0.2880) (250.000,0.8450) (285.714,10.6300) (333.333,7.3880)};
\end{axis}
\end{tikzpicture}\end{minipage}%
\begin{minipage}[b]{0.5\columnwidth}\centering\begin{tikzpicture}[baseline=(current axis.south)]
\begin{axis}[scale only axis, width=2.64cm, height=2.43cm,
  ymode=log, xmin=90, xmax=345, ymin=0.001, ymax=60,
  xtick={100,200,300}, minor xtick={100,150,...,350},
  ytick={0.01,0.1,1,10}, yticklabels={0.01,0.1,1,10},
  major tick length=3pt, minor tick length=1.6pt, tick align=outside,
  tick pos=left, tick label style={font=\scriptsize}, label style={font=\scriptsize},
  xlabel={per-SM rate (kb/s)}, ylabel={},
  ylabel style={yshift=-1ex}, xlabel style={yshift=1ex},
  grid=major, grid style={black!10, line width=.3pt},
  v/.style={blue!80!black, mark=square*, mark size=0.88pt, line width=.6pt},
  a/.style={black, mark=*, mark size=0.88pt, line width=.6pt},
  h/.style={red!80!black, mark=triangle*, mark size=1.12pt, line width=.6pt},
]
\draw[dashed, black!45, line width=.3pt] (axis cs:90,1) -- (axis cs:345,1);
\addplot[v] coordinates {(100.000,0.0100) (200.000,0.0290) (222.222,0.0490) (250.000,0.9330) (285.714,7.5190) (333.333,12.2170)};
\addplot[a] coordinates {(100.000,0.0190) (200.000,0.0100) (222.222,0.0100) (250.000,0.0980) (285.714,3.0570) (333.333,5.8010)};
\addplot[h] coordinates {(100.000,0.0100) (200.000,0.0190) (222.222,0.0290) (250.000,0.0100) (285.714,0.0490) (333.333,0.1850)};
\end{axis}
\end{tikzpicture}\end{minipage}\\[-2pt]
\begin{minipage}[t]{0.5\columnwidth}\centering\footnotesize (c) bank --- per-SM\end{minipage}%
\begin{minipage}[t]{0.5\columnwidth}\centering\footnotesize (d) tag --- per-SM\end{minipage}
\caption{Covert-channel error rate on the V100, A100 and H100, with one sender
and one receiver per SM. (a,b)~per-bit error rate versus aggregate bandwidth;
(c,d)~the per-SM error rate, taken as the median over all SMs, versus the per-SM transmission
rate $1/T$ (one bit per slot of $T$~ns); unlike (a,b), this does not scale with the number of SMs. The dashed line marks a $1\%$ error rate.}
\label{fig:covert}
\end{figure}
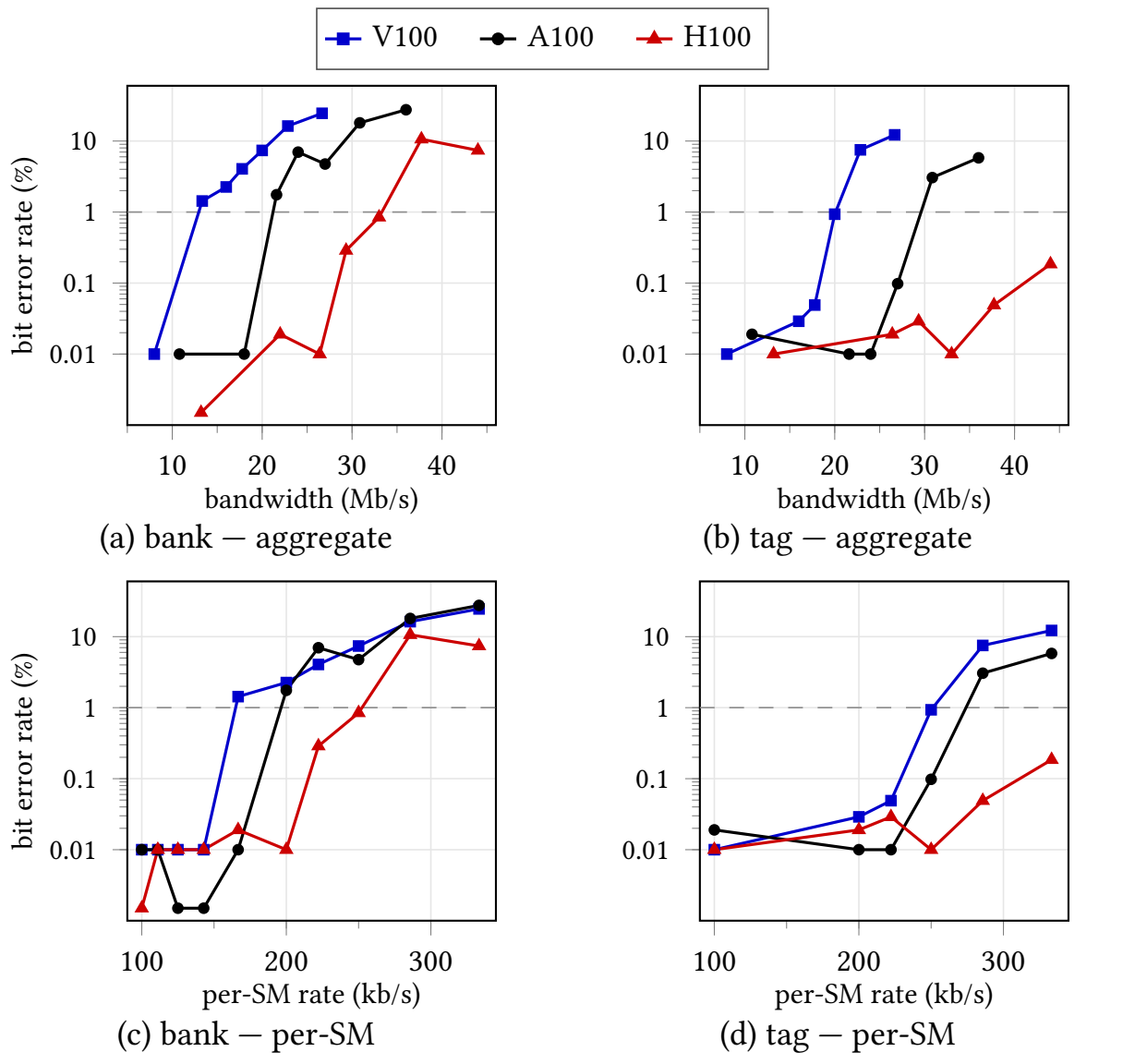

\begin{figure*}[t]
\centering
\begin{minipage}[t]{0.49\textwidth}\centering
\includegraphics[width=\textwidth]{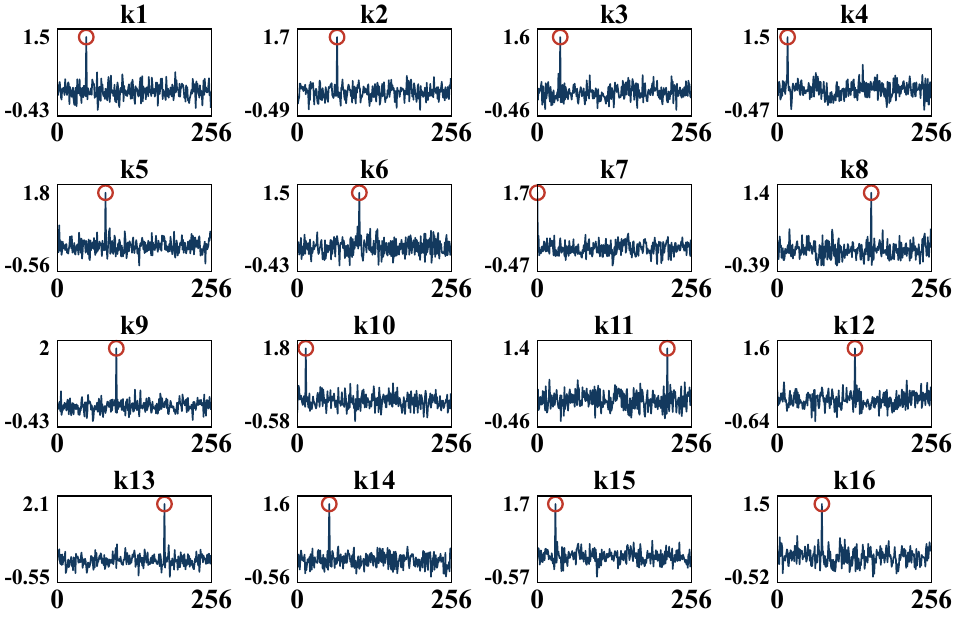}\\[1pt]
{\footnotesize (a) bank channel, standard table, 25M samples}
\end{minipage}\hfill
\begin{minipage}[t]{0.49\textwidth}\centering
\includegraphics[width=\textwidth]{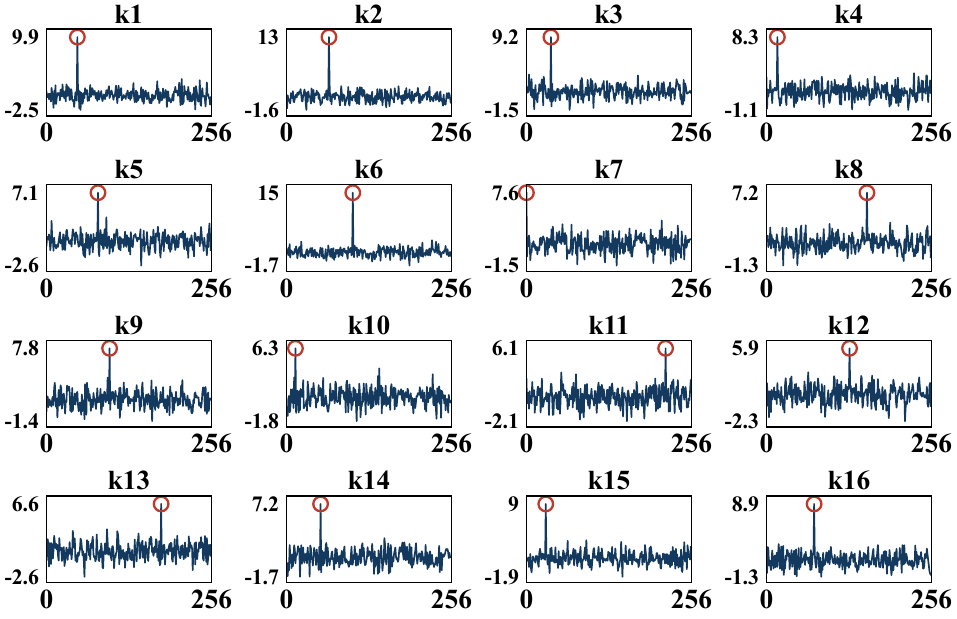}\\[1pt]
{\footnotesize (b) tag channel, replicated table, 1M samples}
\end{minipage}
\caption{AES-128 last-round key recovery on the H100. Each panel is one key byte ($k_1$--$k_{16}$); the horizontal axis is the 256 key-byte guesses and the vertical axis their correlation ($\times10^{-3}$), with the correct guess (circled) at the highest correlation.}
\label{fig:aes}
\end{figure*}

Algorithm~\ref{alg:covert} details our protocol. We use the cross-process GPU
global clock (\texttt{globaltimer}) to divide time into slots of $T$~ns, and
in each slot we transmit one message bit. Because both processes read the same clock, $\mathrm{Synchronization}()$ in
Algorithm~\ref{alg:covert} simply spins each of them until a pre-agreed epoch, without exchanging
any message. To send a ``1'', $\mathrm{contend}()$ runs, on the bank channel, two independent \texttt{LDS}/\texttt{LDG} chains whose loads each take a full bank conflict, saturating the two access channels; on the tag channel it runs a single \texttt{LDG} chain whose loads each incur the maximum tag cost (all $32$ cache lines mapped to one tag unit). To send a ``0'', the sender does nothing, the \emph{do nothing} branch of Algorithm~\ref{alg:covert}. On either channel, $\mathrm{probe}()$ repeatedly times a global load (\texttt{LDG}) chain with no bank conflict and no extra tag cost; the receiver averages the per-slot latencies to suppress noise and raise the success rate, then compares the average against the per-SM median threshold to decide whether the bit is a~0 or a~1.

We implement the protocol in CUDA and compile it with \texttt{nvcc -O3}, without editing the SASS. We evaluate both
channels on the V100, A100, and H100, placing one sender and one receiver process on every SM. Figure~\ref{fig:covert}(a,b) plots the
aggregate bandwidth against the per-bit error rate; shrinking the slot time $T$
raises bandwidth and trades directly against the error rate. On every GPU the
tag channel beats the bank channel, holding an error rate one to two orders of
magnitude lower at the same bandwidth.

\begin{implication}
Both the concurrent bank-access channels and the L1 tag lookup are
exploitable cross-process covert channels, and the tag-lookup channel is the
stronger.
\end{implication}

We measure a peak bandwidth of $44.0$/$27.0$/$20.0$\,Mb/s on the tag channel and
$29.3$/$18.0$/$11.4$\,Mb/s on the bank channel (H100/A100/V100), all at a low
per-bit error rate (below $1\%$); for both channels the usable bandwidth is
highest on the H100, followed by the A100 and then the V100. To trace the
source of this ordering,
Figure~\ref{fig:covert}(c,d) shows, at each per-SM rate $1/T$, the median over all SMs of their
error rates, which therefore does not depend on how many SMs the GPU has; on both channels H100 sustains a lower error rate at a
smaller slot $T$. The same ordering holds per SM: on the bank channel a single
SM keeps its error rate below $1\%$ down to a slot of $7$, $6$ and $4$~$\mu$s on
V100, A100 and H100 respectively, so each generation tolerates a strictly
smaller slot.

\begin{implication}
The usable covert bandwidth of the load path grows with each GPU generation
(H100 $>$ A100 $>$ V100), driven by both more co-located SMs ($132$, $108$ and
$80$) and a smaller slot each SM can tolerate.
\end{implication}

\subsection{Recovering an AES Key through Timing}
\label{sec:impl-sc}

To recover an AES-128 key, we adopt the threat model of prior GPU AES timing
attacks~\cite{jiang16,trident21,jiang19}: a dedicated server offers
AES-128 encryption on its GPU, and an unprivileged attacker repeatedly
submits plaintexts, observes the returned ciphertexts, and obtains the
encryption timing, aiming to recover the 16-byte last-round key. As in
that work, to demonstrate the existence of the channels, our timing is obtained in a clean setting, directly from the
GPU kernel. This is a reasonable setting: as prior work argues, because
the attacker supplies the workload and stress-tests the server, it
controls the configuration and picks the least noisy one
available~\cite{trident21}, and host noise has been shown only to raise
the number of samples needed rather than to defeat the
attack~\cite{jiang16,trident21}; we therefore report our results in the
clean setting. Furthermore, we assume a cold L1, L2 cache, with the
T-tables and all other data in off-chip DRAM.

In \S\ref{sec:machine}, we identify two address-dependent costs for a global load that hits in the L1 cache: the bank
access (\S\ref{sec:bank}) and the tag lookup (\S\ref{sec:tag}). In the last round of table-based AES, the T4 lookup indices
depend on the key, so both costs can become key-dependent timing. We evaluate the bank-conflict
leakage with the standard contiguous 1\,KB T4 table, and then use a bank-conflict-free layout to
examine the leakage along the tag-lookup path.

The second layout follows the table-replication strategy that Tezcan~\cite{tezcan21} uses to
remove shared-memory bank conflicts in GPU AES: the T0 table is duplicated 32 times, one copy per
bank, so that each lane reads its own bank. Since a global load and a shared load resolve bank conflicts by the same mechanism and at the
same cost from the A100 onward (\S\ref{sec:bank}), this shared-memory technique also makes sense
in global memory too. Tezcan replicates only T0 this way, limited by the shared-memory
size of the GPUs of the time, and uses other techniques for the remaining tables. He states that
keeping more replicated tables would yield larger speedups, and on-chip SRAM has grown with each
GPU generation. Hence, with enough on-chip memory, a future implementation could replicate all of
its tables, but here, to demonstrate that the tag channel exists, we replicate only the T4 table,
32 times in global memory, keeping its lookup values while removing intra-warp bank conflicts and
leaving the rest of a standard implementation unchanged. We store the table in a
128-byte-aligned array $R[256][32]$, where $R[i][\ell]=\mathtt{T4}[i]$ and lane $\ell$ reads
$R[i][\ell]$.

For both layouts, we follow the correlation-based last-round attack of Jiang et
al.~\cite{jiang16}. For each key byte we enumerate all 256 candidates and combine each
with the observed ciphertext bytes to infer the lookup index of every lane. We translate
these indices into addresses through the fixed table layout and predict the corresponding cost: for the
standard table, the number of distinct words in the busiest bank (\S\ref{sec:bank}); for the
replicated table, the number of distinct lines on the busiest tag-lookup unit (\S\ref{sec:tag}).
We compute the Pearson correlation between each candidate's predicted costs and the measured
times over all samples, and take the candidate with the highest correlation as the key byte.

We compile both kernels with \texttt{-O3} and run them on the H100~(Fig.~\ref{fig:aes}): the bank
channel recovers all 16 last-round key bytes from the standard table with 25M samples, and the tag channel does
so from the replicated table with 1M samples. We observe the same full key recovery on the A100.
On the V100 we also recover the full key from the tag channel with 8M samples. Even though its global
bank-conflict cost does not follow the A100 and H100 formula, we still recover the full key from the bank
channel: the last encryption round's per-load bank conflicts are small, mostly 1 to 4, where the cost stays roughly linear~(Fig.~\ref{fig:bank}).

\begin{implication}
Bank-conflict leakage is not specific to shared memory; it is also present on
the global-load path.
\end{implication}

\begin{implication}
The L1 tag-lookup units are an independent timing channel.
\end{implication}

\section{Conclusion}

In this paper, we reverse-engineered how the LSU serves loads across the V100, A100, and H100.
We derived cycle-accurate latency formulas for shared loads and, on the A100 and H100, global L1
hits, separating address-dependent bank and tag costs. We also found that LSU saturation can
nearly starve later-launched blocks. These findings enabled full AES-128 key recovery through two
global-load timing channels and cross-process covert communication under MPS, while one warp
slowed cuBLAS GEMM by up to 563$\times$ on the H100.

\bibliographystyle{ACM-Reference-Format}
\bibliography{refs_acm}

@misc{cuassembler,
  author       = {{cloudcores}},
  title        = {{CuAssembler: An Unofficial CUDA Assembler}},
  howpublished = {\url{https://github.com/cloudcores/CuAssembler}},
  year         = {2021}
}

@misc{a100wp,
  author       = {{NVIDIA}},
  title        = {{NVIDIA A100 Tensor Core GPU Architecture}},
  howpublished = {Whitepaper},
  year         = {2020}
}

@misc{h100wp,
  author       = {{NVIDIA}},
  title        = {{NVIDIA H100 Tensor Core GPU Architecture}},
  howpublished = {Whitepaper},
  year         = {2022}
}

@inproceedings{huerta25,
  author    = {Huerta, Rodrigo and Abaie Shoushtary, Mojtaba and Cruz, Jos\'{e}-Lorenzo and Gonz\'{a}lez, Antonio},
  title     = {{Dissecting and Modeling the Architecture of Modern GPU Cores}},
  booktitle = {Proceedings of the 58th IEEE/ACM International Symposium on Microarchitecture (MICRO)},
  year      = {2025}
}

@misc{jia18,
  author = {Jia, Zhe and Maggioni, Marco and Staiger, Benjamin and Scarpazza, Daniele P.},
  title  = {{Dissecting the NVIDIA Volta GPU Architecture via Microbenchmarking}},
  note   = {arXiv:1804.06826},
  year   = {2018}
}

@misc{jia19,
  author = {Jia, Zhe and Maggioni, Marco and Smith, Jeffrey and Scarpazza, Daniele P.},
  title  = {{Dissecting the NVidia Turing T4 GPU via Microbenchmarking}},
  note   = {arXiv:1903.07486},
  year   = {2019}
}

@article{mei16,
  author  = {Mei, Xinxin and Chu, Xiaowen},
  title   = {{Dissecting GPU Memory Hierarchy Through Microbenchmarking}},
  journal = {IEEE Transactions on Parallel and Distributed Systems},
  volume  = {28},
  number  = {1},
  pages   = {72--86},
  year    = {2017}
}

@article{sun22,
  author  = {Sun, Wei and Li, Ang and Geng, Tong and Stuijk, Sander and Corporaal, Henk},
  title   = {{Dissecting Tensor Cores via Microbenchmarks: Latency, Throughput and Numeric Behaviors}},
  journal = {IEEE Transactions on Parallel and Distributed Systems},
  volume  = {34},
  number  = {1},
  pages   = {246--261},
  year    = {2023}
}

@article{jiang19,
  author  = {Jiang, Zhen Hang and Fei, Yunsi and Kaeli, David},
  title   = {{Exploiting Bank Conflict-based Side-channel Timing Leakage of GPUs}},
  journal = {ACM Transactions on Architecture and Code Optimization},
  volume  = {16},
  number  = {4},
  articleno = {42},
  year    = {2019}
}

@inproceedings{nugteren14,
  author    = {Nugteren, Cedric and van den Braak, Gert-Jan and Corporaal, Henk and Bal, Henri},
  title     = {{A Detailed GPU Cache Model Based on Reuse Distance Theory}},
  booktitle = {Proceedings of the 20th IEEE International Symposium on High Performance Computer Architecture (HPCA)},
  pages     = {37--48},
  year      = {2014}
}

@inproceedings{irazoqui15,
  author    = {Irazoqui, Gorka and Eisenbarth, Thomas and Sunar, Berk},
  title     = {{Systematic Reverse Engineering of Cache Slice Selection in Intel Processors}},
  booktitle = {Proceedings of the Euromicro Conference on Digital System Design (DSD)},
  year      = {2015}
}

@inproceedings{knockknock25,
  author    = {Plin, Antoine and Casalino, Lorenzo and Rokicki, Thomas and Salvador, Ruben},
  title     = {{Knock-Knock: Black-Box, Platform-Agnostic DRAM Address-Mapping Reverse Engineering}},
  booktitle = {Proceedings of the 2nd Microarchitecture Security Conference (uASC '26)},
  year      = {2026},
  note      = {arXiv:2509.19568}
}

@inproceedings{jain19,
  author    = {Jain, Saksham and Baek, Iljoo and Wang, Shige and Rajkumar, Ragunathan},
  title     = {{Fractional GPUs: Software-Based Compute and Memory Bandwidth Reservation for GPUs}},
  booktitle = {Proceedings of the IEEE Real-Time and Embedded Technology and Applications Symposium (RTAS)},
  year      = {2019}
}

@misc{v100wp,
  author       = {{NVIDIA}},
  title        = {{NVIDIA Tesla V100 GPU Architecture}},
  howpublished = {Whitepaper},
  year         = {2017}
}

@book{aamodt18,
  author    = {Aamodt, Tor M. and Fung, Wilson Wai Lun and Rogers, Timothy G.},
  title     = {{General-Purpose Graphics Processor Architectures}},
  series    = {Synthesis Lectures on Computer Architecture},
  publisher = {Morgan \& Claypool},
  year      = {2018}
}

@misc{nvforum26,
  author       = {{NVIDIA Developer Forums}},
  title        = {{Nsight Compute seems to be giving the wrong number of bank conflicts}},
  howpublished = {\url{https://forums.developer.nvidia.com/t/nsight-compute-seems-to-be-giving-the-wrong-number-of-bank-conflicts/363034}},
  note         = {Reply by Greg, Mar.~11, 2026},
  year         = {2026}
}

@misc{nvforum25tag,
  author       = {{NVIDIA Developer Forums}},
  title        = {{Question about L1 tag stage resolution per cycle}},
  howpublished = {\url{https://forums.developer.nvidia.com/t/question-about-l1-tag-stage-resolution-per-cycle/354027}},
  note         = {Reply by Greg, Dec.~9, 2025},
  year         = {2025}
}

@inproceedings{zhang17,
  author    = {Zhang, Xiuxia and Tan, Guangming and Xue, Shuangbai and Li, Jiajia and Zhou, Keren and Chen, Mingyu},
  title     = {{Understanding the GPU Microarchitecture to Achieve Bare-Metal Performance Tuning}},
  booktitle = {Proceedings of the 22nd ACM SIGPLAN Symposium on Principles and Practice of Parallel Programming (PPoPP)},
  year      = {2017}
}

@inproceedings{raihan19,
  author    = {Raihan, Md Aamir and Goli, Negar and Aamodt, Tor M.},
  title     = {{Modeling Deep Learning Accelerator Enabled GPUs}},
  booktitle = {Proceedings of the IEEE International Symposium on Performance Analysis of Systems and Software (ISPASS)},
  year      = {2019}
}

@misc{cudabp,
  author       = {{NVIDIA}},
  title        = {{CUDA C++ Best Practices Guide}},
  howpublished = {\url{https://docs.nvidia.com/cuda/cuda-c-best-practices-guide/}},
  note         = {Section 10.2.3.1, Shared Memory and Memory Banks; accessed Oct.~2026},
  year         = {2026}
}

@misc{nvpat9348762,
  author       = {Fahs, Brian and Anderson, Eric T. and Barrow-Williams, Nick and Gadre, Shirish and McCormack, Joel James and Nordquist, Bryon S. and Saxena, Nirmal Raj and Shah, Lacky V.},
  title        = {{Technique for accessing content-addressable memory}},
  howpublished = {U.S. Patent 9,348,762 B2, NVIDIA Corporation},
  note         = {Filed Dec.~19, 2012; granted May~24, 2016},
  year         = {2016}
}

@misc{nvpat11080051,
  author       = {Kerr, Andrew and Choquette, Jack and Qiu, Xiaogang and Paranjape, Omkar and Rao, Poornachandra and Gadre, Shirish and Heinrich, Steven J. and Patel, Manan and Giroux, Olivier and Kaatz, Alan},
  title        = {{Techniques for efficiently transferring data to a processor}},
  howpublished = {U.S. Patent 11,080,051 B2, NVIDIA Corporation},
  note         = {Priority Oct.~29, 2019; granted Aug.~3, 2021},
  year         = {2021}
}

@misc{keplerwp,
  author       = {{NVIDIA}},
  title        = {{NVIDIA's Next Generation CUDA Compute Architecture: Kepler GK110/210}},
  howpublished = {Whitepaper, v1.1},
  year         = {2014}
}

@inproceedings{jiang16,
  title={A Complete Key Recovery Timing Attack on a {GPU}},
  author={Jiang, Zhen Hang and Fei, Yunsi and Kaeli, David},
  booktitle={Proc. IEEE Int. Symp. High Performance Computer Architecture (HPCA)},
  year={2016}
}

@inproceedings{trident21,
  title={Trident: A Hybrid Correlation-Collision {GPU} Cache Timing Attack for {AES} Key Recovery},
  author={Ahn, Jaeguk and Jin, Cheolgyu and Kim, Jiho and Rhu, Minsoo and Fei, Yunsi and Kaeli, David and Kim, John},
  booktitle={Proc. IEEE Int. Symp. High Performance Computer Architecture (HPCA)},
  year={2021}
}

@inproceedings{naghib17,
  title={Constructing and Characterizing Covert Channels on {GPGPUs}},
  author={Naghibijouybari, Hoda and Khasawneh, Khaled N. and Abu-Ghazaleh, Nael},
  booktitle={Proc. IEEE/ACM Int. Symp. Microarchitecture (MICRO)},
  year={2017}
}

@inproceedings{ahnnoc21,
  title={Network-on-Chip Microarchitecture-based Covert Channel in {GPUs}},
  author={Ahn, Jaeguk and Kim, Jiho and Kasan, Hans and Delshadtehrani, Leila and Song, Wonjun and Joshi, Ajay and Kim, John},
  booktitle={Proc. IEEE/ACM Int. Symp. Microarchitecture (MICRO)},
  year={2021}
}

@inproceedings{dutta23,
  title={Spy in the {GPU}-box: Covert and Side Channel Attacks on Multi-{GPU} Systems},
  author={Dutta, Sankha Baran and Naghibijouybari, Hoda and Gupta, Arjun and Abu-Ghazaleh, Nael and Marquez, Andres and Barker, Kevin},
  booktitle={Proc. Int. Symp. Computer Architecture (ISCA)},
  year={2023}
}

@inproceedings{tunnels23,
  title={{TunneLs} for Bootlegging: Fully Reverse-Engineering {GPU} {TLBs} for Challenging Isolation Guarantees of {NVIDIA} {MIG}},
  author={Zhang, Zhenkai and Allen, Tyler and Yao, Fan and Gao, Xing and Ge, Rong},
  booktitle={Proc. ACM SIGSAC Conf. Computer and Communications Security (CCS)},
  year={2023}
}

@inproceedings{wang25,
  title={Security and Performance Implications of {GPU} Cache Eviction Priority Hints},
  author={Wang, Qizhong and Huang, Xiangyue and Guo, Yanan and Xu, Yuanchao},
  booktitle={Proc. IEEE/ACM Int. Symp. Microarchitecture (MICRO)},
  year={2025}
}

@inproceedings{nayak21,
  title={{(Mis)managed}: A Novel {TLB}-based Covert Channel on {GPUs}},
  author={Nayak, Ajay and {Pratheek B.} and Ganapathy, Vinod and Basu, Arkaprava},
  booktitle={Proc. ACM Asia Conf. Computer and Communications Security (ASIA CCS)},
  year={2021}
}

@inproceedings{miao24,
  title={Veiled Pathways: Investigating Covert and Side Channels within {GPU} Uncore},
  author={Miao, Yuanqing and Zhang, Yingtian and Wu, Dinghao and Zhang, Danfeng and Tan, Gang and Zhang, Rui and Kandemir, Mahmut Taylan},
  booktitle={Proc. IEEE/ACM Int. Symp. Microarchitecture (MICRO)},
  year={2024}
}

@inproceedings{mc3,
  title={{MC3}: Memory Contention-based Covert Channel Communication on Shared {DRAM} System-on-Chips},
  author={Dagli, Ismet and Crea, James and Seckiner, Soner and Xu, Yuanchao and K{\"o}se, Sel{\c{c}}uk and Belviranli, Mehmet E.},
  booktitle={Proc. Design, Automation \& Test in Europe (DATE)},
  year={2025}
}

@inproceedings{llc15,
  title={Last-Level Cache Side-Channel Attacks are Practical},
  author={Liu, Fangfei and Yarom, Yuval and Ge, Qian and Heiser, Gernot and Lee, Ruby B.},
  booktitle={Proc. IEEE Symp. Security and Privacy (S\&P)},
  year={2015}
}

@inproceedings{yandrofski22,
  author={Yandrofski, Tyler and Chen, Jingyuan and Otterness, Nathan and Anderson, James H. and Smith, F. Donelson},
  title={{Making Powerful Enemies on NVIDIA GPUs}},
  booktitle={Proc. IEEE Real-Time Systems Symposium (RTSS)},
  pages={383--395},
  year={2022}
}

@misc{ncuguide,
  author       = {{NVIDIA}},
  title        = {{Nsight Compute Kernel Profiling Guide}},
  howpublished = {\url{https://docs.nvidia.com/nsight-compute/ProfilingGuide/}},
  note         = {Version 2026.3.1, Section 2.4.7, Warp Stall Reasons; accessed Oct.~2026},
  year         = {2026}
}

@misc{cutlass,
  author       = {{NVIDIA}},
  title        = {{CUTLASS: CUDA Templates for Linear Algebra Subroutines}},
  howpublished = {\url{https://github.com/NVIDIA/cutlass}},
  note         = {\texttt{media/docs/cpp/efficient\_gemm.md}; accessed Oct.~2026},
  year         = {2026}
}

@inproceedings{dao22,
  title={{FlashAttention}: Fast and Memory-Efficient Exact Attention with {IO}-Awareness},
  author={Dao, Tri and Fu, Daniel Y. and Ermon, Stefano and Rudra, Atri and R\'{e}, Christopher},
  booktitle={Advances in Neural Information Processing Systems (NeurIPS)},
  year={2022}
}

@inproceedings{dao23,
  title={{FlashAttention-2}: Faster Attention with Better Parallelism and Work Partitioning},
  author={Dao, Tri},
  booktitle={Proc. Int. Conf. Learning Representations (ICLR)},
  year={2024}
}

@inproceedings{shah24,
  title={{FlashAttention-3}: Fast and Accurate Attention with Asynchrony and Low-precision},
  author={Shah, Jay and Bikshandi, Ganesh and Zhang, Ying and Thakkar, Vijay and Ramani, Pradeep and Dao, Tri},
  booktitle={Advances in Neural Information Processing Systems (NeurIPS)},
  year={2024}
}

@inproceedings{kadam18,
  title={{RCoal}: Mitigating {GPU} Timing Attack via Subwarp-based Randomized Coalescing Techniques},
  author={Kadam, Gurunath and Zhang, Danfeng and Jog, Adwait},
  booktitle={Proc. IEEE Int. Symp. High Performance Computer Architecture (HPCA)},
  year={2018}
}

@inproceedings{kadam20,
  title={{BCoal}: Bucketing-based Memory Coalescing for Efficient and Secure {GPUs}},
  author={Kadam, Gurunath and Zhang, Danfeng and Jog, Adwait},
  booktitle={Proc. IEEE Int. Symp. High Performance Computer Architecture (HPCA)},
  year={2020}
}

@inproceedings{xu19,
  title={{GPUGuard}: Mitigating Contention Based Side and Covert Channel Attacks on {GPUs}},
  author={Xu, Qiumin and Naghibijouybari, Hoda and Wang, Shibo and Abu-Ghazaleh, Nael and Annavaram, Murali},
  booktitle={Proc. ACM Int. Conf. Supercomputing (ICS)},
  year={2019}
}

@inproceedings{megatron21,
  author    = {Narayanan, Deepak and Shoeybi, Mohammad and Casper, Jared and LeGresley, Patrick and Patwary, Mostofa and Korthikanti, Vijay and Vainbrand, Dmitri and Kashinkunti, Prethvi and Bernauer, Julie and Catanzaro, Bryan and Phanishayee, Amar and Zaharia, Matei},
  title     = {{Efficient Large-Scale Language Model Training on GPU Clusters Using Megatron-LM}},
  booktitle = {Proceedings of the International Conference for High Performance Computing, Networking, Storage and Analysis (SC)},
  year      = {2021},
  pages     = {1--15},
  doi       = {10.1145/3458817.3476209}
}

@misc{top500jun26,
  author       = {{TOP500}},
  title        = {{TOP500 List, June 2026}},
  howpublished = {\url{https://top500.org/lists/top500/2026/06/}},
  note         = {67th edition; eight of the ten fastest systems use GPUs (AMD Instinct, Intel Data Center GPU Max, NVIDIA H100/GH200); accessed Nov.~2026},
  year         = {2026}
}

@article{tezcan21,
  author  = {Tezcan, Cihangir},
  title   = {{Optimization of Advanced Encryption Standard on Graphics Processing Units}},
  journal = {IEEE Access},
  volume  = {9},
  pages   = {67315--67326},
  year    = {2021},
  doi     = {10.1109/ACCESS.2021.3077551}
}

@inproceedings{bechtel22,
  author={Bechtel, Michael and Yun, Heechul},
  title={{Denial-of-Service Attacks on Shared Resources in Intel's Integrated CPU-GPU Platforms}},
  booktitle={Proc. IEEE Int. Symp. Real-Time Distributed Computing (ISORC)},
  pages={1--10},
  year={2022},
  doi={10.1109/ISORC52572.2022.9812711}
}

@inproceedings{olmedo20,
  author={Olmedo, Ignacio Sa\~{n}udo and Capodieci, Nicola and Martinez, Jorge Luis and Marongiu, Andrea and Bertogna, Marko},
  title={{Dissecting the CUDA Scheduling Hierarchy: A Performance and Predictability Perspective}},
  booktitle={Proc. IEEE Real-Time and Embedded Technology and Applications Symp. (RTAS)},
  pages={213--225},
  year={2020},
  doi={10.1109/RTAS48715.2020.000-5}
}

@inproceedings{iwai10,
  author={Iwai, Keisuke and Kurokawa, Takakazu and Nishikawa, Naoki},
  title={{AES Encryption Implementation on CUDA GPU and Its Analysis}},
  booktitle={Proc. Int. Conf. Networking and Computing (ICNC)},
  pages={209--214},
  year={2010},
  doi={10.1109/IC-NC.2010.49}
}

\end{document}